\documentclass[twoside,twocolumn,english,aps,showpacs,prl,superscriptaddress,longbibliography]{revtex4-2}
\usepackage[T1]{fontenc}
\usepackage{color}
\usepackage{bm}
\usepackage{amsmath}
\usepackage{amsthm}
\usepackage{amssymb}
\usepackage{graphicx}
\usepackage{subfigure}
\usepackage{esint}
\usepackage{mathrsfs}
\usepackage{booktabs}
\usepackage{multirow}
\usepackage{makecell}
\usepackage{verbatim}
\usepackage{physics}
\usepackage[plainpages = false, pdfpagelabels, bookmarksnumbered = true,
                 breaklinks = true,
                 linktocpage,
                 colorlinks = true,
                 linkcolor = blue,
                 urlcolor  = blue,
                 citecolor = blue,
                 anchorcolor = green,
                 hyperindex = true,
                 hyperfigures
                 ]{hyperref} 
\usepackage{dcolumn}

\begin{document}

\preprint{APS/123-QED}

\title{Tip-Tuned Renormalization-Group Spectroscopy Unmasks a False-positive Topological Superconducting Vortex}

\author{Zhenhua Zhu}
\thanks{These authors contributed equally to this work.}
\affiliation{State Key Laboratory of Low Dimensional Quantum Physics, Department of Physics, Tsinghua University, Beijing, 100084, China}
\affiliation{Frontier Science Center for Quantum Information, Beijing 100184, China}
\author{Qun Zhu}
\thanks{These authors contributed equally to this work.}
\affiliation{State Key Laboratory of Low Dimensional Quantum Physics, Department of Physics, Tsinghua University, Beijing, 100084, China}
\author{Yong-Wei Wang}
\affiliation{State Key Laboratory of Low Dimensional Quantum Physics, Department of Physics, Tsinghua University, Beijing, 100084, China}
\author{Gu Zhang}
\affiliation{National Laboratory of Solid State Microstructures, School of Physics, Jiangsu Physical Science Research Center and Collaborative Innovation Center of Advanced Microstructures, Nanjing University, Nanjing 210093, China}
\author{Jihai Zhang}
\affiliation{State Key Laboratory of Low Dimensional Quantum Physics, Department of Physics, Tsinghua University, Beijing, 100084, China}
\author{Xu-Cun Ma}
\affiliation{State Key Laboratory of Low Dimensional Quantum Physics, Department of Physics, Tsinghua University, Beijing, 100084, China}
\affiliation{Frontier Science Center for Quantum Information, Beijing 100184, China}
\author{Qi-Kun Xue}
\affiliation{State Key Laboratory of Low Dimensional Quantum Physics, Department of Physics, Tsinghua University, Beijing, 100084, China}
\affiliation{Frontier Science Center for Quantum Information, Beijing 100184, China}
\affiliation{Shenzhen Institute for Quantum Science and Engineering and Department of Physics, Southern University of Science and Technology, Shenzhen 518055, China}
\affiliation{Beijing Academy of Quantum Information Sciences, Beijing 100193, China}
\affiliation{Hefei National Laboratory, Hefei 230088, China}
\author{Can-Li Song}
\email[Corresponding to: ]{clsong07@mail.tsinghua.edu.cn}
\affiliation{State Key Laboratory of Low Dimensional Quantum Physics, Department of Physics, Tsinghua University, Beijing, 100084, China}
\affiliation{Frontier Science Center for Quantum Information, Beijing 100184, China}
\author{Dong E. Liu}
\email[Corresponding to: ]{dongeliu@mail.tsinghua.edu.cn}
\affiliation{State Key Laboratory of Low Dimensional Quantum Physics, Department of Physics, Tsinghua University, Beijing, 100084, China}
\affiliation{Frontier Science Center for Quantum Information, Beijing 100184, China}
\affiliation{Beijing Academy of Quantum Information Sciences, Beijing 100193, China}
\affiliation{Hefei National Laboratory, Hefei 230088, China}

\begin{abstract}
Clean, nonsplit vortex zero-bias peaks (ZBPs) can be misinterpreted as Majorana zero modes (MZMs), making static scanning tunneling microscopy intrinsically ambiguous. Here we use the STM tip coupling to drive a local boundary-renormalization-group (boundary RG) flow, turning dynamical Coulomb blockade into a falsification test for Majorana-like ZBPs. Experimentally, in a $\mathrm{SrSn}_3$ thin film, normal-state spectra establish an Ohmic dissipative environment, and a common boundary-RG/thermodynamic-Bethe-ansatz analysis of the superconducting-gap and vortex-center spectra yields consistent dissipation strengths within the $r < 1/2$ Majorana-filter regime. Lowering the tip nevertheless drives a clean, non-split vortex-center ZBP into a zero-bias dip, opposite to the protected flow of an isolated MZM, unmasking the peak as a Majorana false positive produced by a conventional vortex-core state. The same flow selectively suppresses the strongly tip-coupled channel, resolving the two-gap superconductivity. Dissipative STM thus tests dynamical protection rather than spectral appearance.
\end{abstract}

\maketitle

\section{Introduction}
A central bottleneck in the search for Majorana zero modes (MZMs)~\cite{readgreen2000paired,kitaev2001unpaired,alicea2012directions} is the intrinsic ambiguity of zero-bias conductance peaks (ZBPs)~\cite{flensberg2010tunneling,law2009majorana}. Andreev bound states, impurity resonances, unresolved Caroli--de Gennes--Matricon (CdGM) levels, and other non-topological mechanisms can produce closely similar low-energy features~\cite{liu2012zbp,kells2012trivial,prada2012transport,pan2020physical_zbp,yu2021nonmajorana,pan2021quantized_unquantized,valentini2021nontopological,kim2021anisotropic}. This challenge is acute in scanning tunneling microscopy/spectroscopy (STM/STS) of superconducting vortices, where $dI/dV$ probes the local density of states projected onto the tip orbital~\cite{caroli1964vortex,binnig1982controllable_gap,tersoff1983theory_application,hess1989vortex,pan2000vortex,kawakami2015evolution}. Milestone vortex experiments reporting clean, non-split vortex-center ZBPs have established STM as a compelling spectroscopic platform for potential MZMs~\cite{fu2008proximity,wang2018fetse_majorana,liu2018robust,machida2019zero_energy,kong2019half,zhu2020quantized,song2026charge_stripes_mzm}. The key experimental frontier is to conﬁrm the topological protection absent in trivial systems.

\begin{figure}[t]
    \centering
    \includegraphics[width=0.9\linewidth]{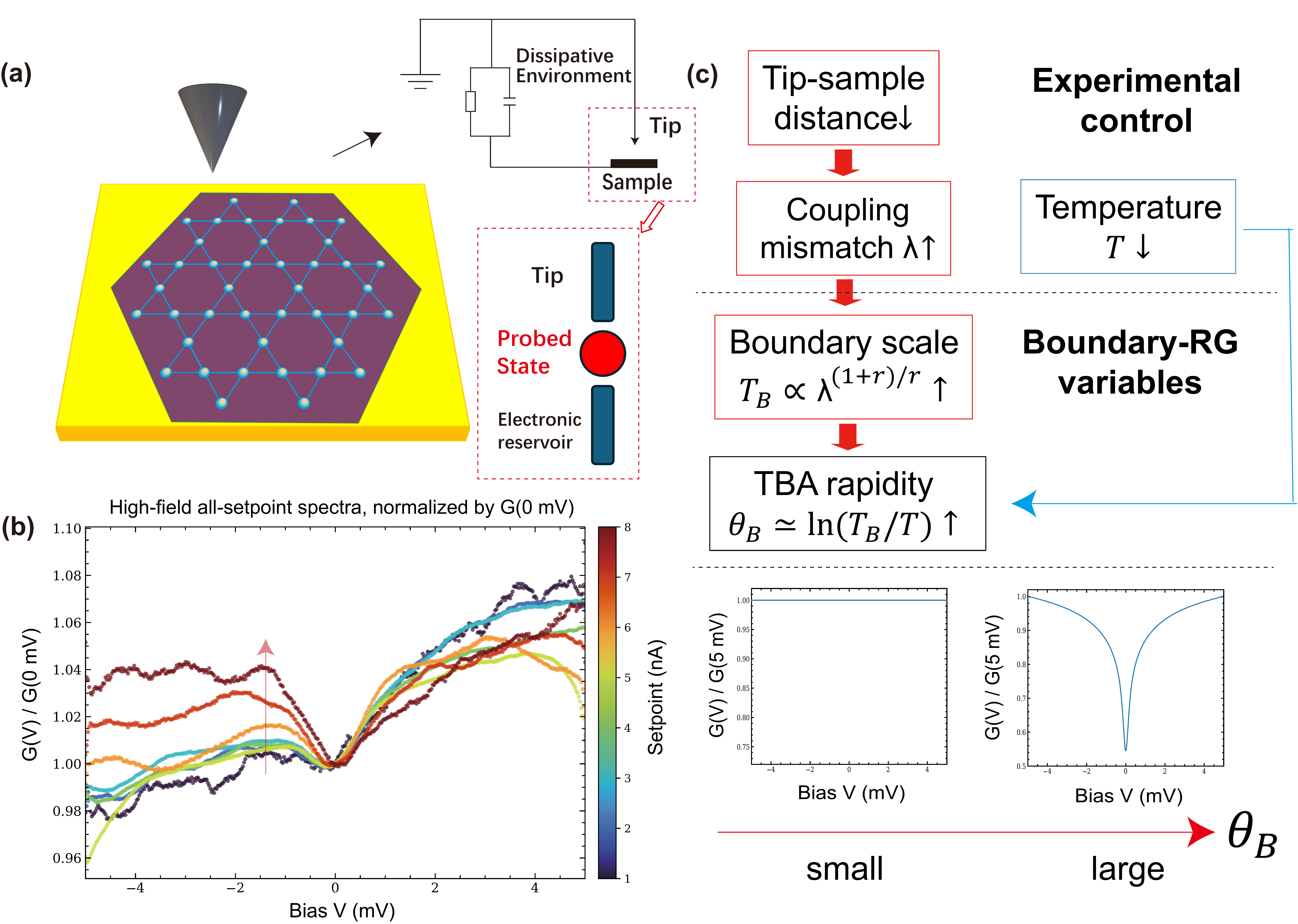}
    \caption{
    Dissipative STM junction and normal-state DCB response.
    \textbf{(a)} Illustration of setup, where a tip is detecting the sample (blue) on a substrate (yellow), with a
    disorder layer between them (represented by purple).
    \textbf{Insets}: (Upper) Effective circuit of STM junction in dissipative environment. (Lower) An illustration circuit of a lead-dot-lead (blue-red-blue) device. 
    \textbf{(b)} $1$ T spectra after superconductivity is
    suppressed, showing a zero-bias anomaly that strengthens as the tip is
    lowered (highlighted by red arrow). The fitted dissipation strength is $r_{\rm HF}\approx 0.22$.
    \textbf{(c)} Logic flow for the effect of lowering tip in RG theory. The right panels plot representative DCB response of a normal metallic DOS. An Ohmic environment suppresses low-energy tunneling and produces a
    zero-bias conductance valley when the tip is lowering.
    }
    \label{fig:setup_dcb}
\end{figure}

Dynamical Coulomb blockade (DCB) provides the physical foundation for this test. Coupling an STM junction to an Ohmic electromagnetic environment turns it from a passive density-of-states probe into a dissipative quantum circuit~\cite{caldeira1983quantum,schmid1983diffusion,devoret1990electromagnetic,girvin1990quantum,ingold1992charge,safi2004one_channel,parmentier2011backaction,brun2012dcb_contacts,mebrahtu2013majorana_critical,senkpiel2020local_probe,zhang2021conductance,liu2024dynamical}, whose backaction renormalizes the low-bias conductance. This quantum transport is then governed by the dissipation strength $r=R/R_Q$ and the bare tunneling amplitude, which STM tunes continuously through the tip-sample distance. DCB thereby realizes the ``Majorana filter'' proposed by Ref.~\cite{liu2013majorana_filter}: an MZM-induced zero-bias peak survives the DCB suppression due to the topological protection, whereas a trivial zero-energy resonance, incontrast, is driven into a zero-bias dip as probe coupling increases. This principle has been extended to Andreev reflection~\cite{liu2022universal} and vortex STM~\cite{zhang2023vortex_dissipative}, and explored in hybrid nanowire devices~\cite{zhang2022dissipative_lead,wang2022weak_dissipation,zhang2023insitu_dcb}.

Here we establish tip-height-dependent STM as an experimental boundary-renormalization-group (boundary-RG) spectroscopy. The probed state, coupled to the STM tip on one side and to the electronic continuum of the sample on the other, forms an effective lead–dot–lead device whose transmission is maximal when its hybridizations with the two reservoirs are matched~\cite{zhang2021conductance}. For the overcoupled dominant channel studied here, lowering the tip enlarges this coupling mismatch, raises the boundary crossover scale, and drives the spectrum along a controlled RG trajectory, which is a sharper diagnostic than any fixed-coupling ZBP. The boundary sine-Gordon/thermodynamic Bethe ansatz (TBA) solution~\cite{fendley1995exact,boulat2019full} captures the resulting DCB-modified transport nonperturbatively, beyond the perturbative $P(E)$ treatment~\cite{ingold1992charge,serrier2013scanning}, converting the dissipative vortex-STM criterion~\cite{zhang2023vortex_dissipative} into a calibrated boundary-RG spectroscopy.

\begin{figure*}[t]
    \centering
    \includegraphics[width=\linewidth]{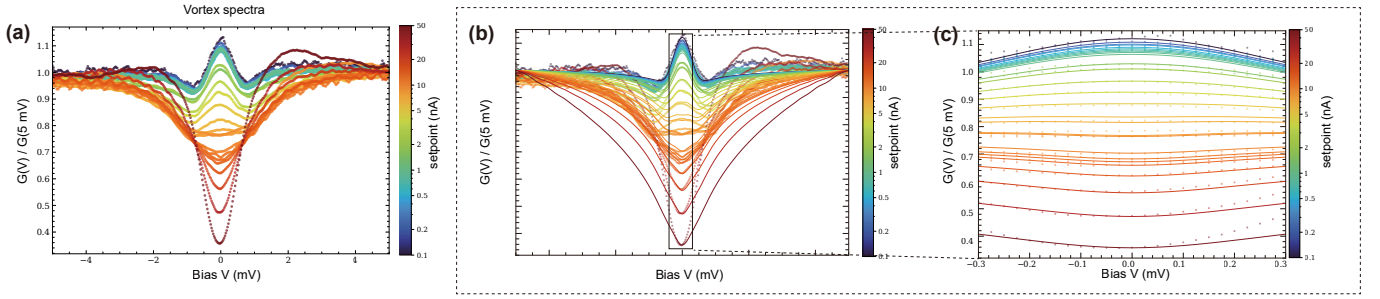}
    \caption{
    \textbf{Single-channel boundary-RG analysis of vortex-center spectra.}
    \textbf{(a)} Experimental vortex-center spectra on the $37.6$-nm island, showing a clean non-split zero-bias peak at low setpoint that evolves into a zero-bias dip as the tip is lowered. The color scale denotes the setpoint current.
    \textbf{(b)} Single-continuum-channel DCB/TBA fit to the vortex-center spectra with Eq.~\eqref{dressed_conductance}, giving $r_{\rm v}^s\approx 0.19$.
    \textbf{(c)} Low-bias zoom of (b), showing that the model captures the central peak-to-dip boundary flow. Dots are experimental data and solid curves are fits.
    }
\label{fig:sc_vortex_single_channel}
\end{figure*}

We implement this framework in SrSn$_3$ thin films, a superconductor reported to host two-component superconductivity and topological bands~\cite{song2025srsn3,yang2025srsn3}. Normal-state spectra identify an Ohmic environment, and TBA fits to the superconducting-gap and vortex-center spectra yield consistent dissipation strengths, $r_{\rm SC}\simeq0.21$ and $r_{\rm v}\simeq0.18$, placing the junction in the $r<1/2$ regime where the Majorana filter operates.~\cite{liu2013majorana_filter}. The central experimental observation is that a clean non-split vortex-center ZBP, a canonical Majorana candidate in static STM, is continuously suppressed into a zero-bias dip as the tip is lowered, opposite to the protected flow of an isolated MZM and quantitatively captured by our two-gap TBA analysis. The observed ZBP is thus a non-topological vortex-core feature: a concrete Majorana false positive. The same boundary-RG flow further resolves the two-gap superconductivity by selectively suppressing the strongly tip-coupled channel, establishing dissipative STM as both a dynamical Majorana filter and a channel-selective probe.

\section{General setup and DCB}
As illustrated in Fig.~\ref{fig:setup_dcb}a, the STM junction is modeled as a vacuum tunneling barrier embedded within a local Ohmic environment, characterized by an effective resistance $R$ and capacitance at low frequencies~\cite{devoret1990electromagnetic,girvin1990quantum,ingold1992charge}. For single-electron tunneling, we define the dimensionless dissipation strength $r = R/R_Q$, where $R_Q = h/e^2$.
Environmental phase fluctuations renormalize low-energy transport.
In the weak-tunneling limit, the probability of the environment absorbing energy $E$ scales as $P(E) \propto E^{2r -1}\Theta(E)$~\cite{devoret1990electromagnetic,girvin1990quantum,ingold1992charge}. Consequently, tunneling between normal metals with a featureless density of states exhibits a DCB-induced zero-bias dip, $G(V, T=0) \propto |eV|^{2r}$.

Crucially, this dissipative environment differentiates a Majorana mode from an ordinary resonance. Following the Majorana-filter analysis~\cite{liu2013majorana_filter,liu2022universal,zhang2023vortex_dissipative}, an isolated MZM in the $r < 1/2$ regime flows toward a protected fixed point instead of the ordinary DCB valley. In the strong-coupling limit, its zero-bias conductance approaches~\cite{liu2013majorana_filter}
\begin{equation}
\frac{2e^2}{h}-G_{\rm MZM}(V=0,T)
\propto
T^{\frac{2-4r}{1+2r}}.
\label{eq:mzm_conductance}
\end{equation}
By contrast, an ordinary vortex-core resonance remains part of the local sample spectral density, dressed by environmental low-energy suppression. This dichotomy forms the topological filtering principle of our experiment.

Though in Eq.~\eqref{eq:mzm_conductance} the variable is temperature, according to Refs.~\cite{fendley1995exact,boulat2019full,zhang2021conductance,zhang2023vortex_dissipative}, the conductance is dependent on $T_B(\lambda_0)/T$ in boundary-RG. At fixed temperature, increasing $T_B/T$ drives the junction along the same low-energy RG trajectory as lowering the temperature. While lowering the tip increases the relevant Majorana hybridization scale and decreases $T_B^M$~\cite{zhang2023vortex_dissipative}, we will show that it increases the mismatch scale $T_B(\lambda_0)$ for an overcoupled conventional resonance (which will be defined later), producing opposite low-energy flows. 

To describe this crossover, we treat the probed state, the STM tip, and the electronic continuum of the sample as an effective lead-dot-lead device, as illustrated in the lower inset of Fig.~\ref{fig:setup_dcb}(a). The backscattering amplitude $\lambda_0$ measures the departure from the hybridization-matched transmission point (see details in App.~\ref{app_TBA}). For the dominant channel studied here, the observed flow is consistent with the overcoupled regime $\Gamma_t>\Gamma_e$. Lowering the tip then increases $\Gamma_t$, enlarges the coupling mismatch, and consequently increases both $\lambda_0$ and $T_B$. An ordinary zero energy resonance is therefore driven toward stronger low-energy suppression, whereas an isolated MZM in the $r<1/2$ regime flows toward its protected strong-coupling fixed point. The direction of the tip-height evolution thus provides a dynamical distinction that is absent from a single static spectrum.

\section{Boundary scale and TBA response}
The perturbative $P(E)$ description is insufficient once the tip--sample coupling is no longer weak.
The dissipative tunneling problem is naturally connected to the boundary-impurity and resonant-tunneling RG framework.
The nonperturbative TBA solution of the corresponding boundary sine-Gordon model gives a universal DCB scaling function,
$G_{r}[eV/T,T_B/T]$, for a structureless tunneling channel. For fixed dissipation strength $r$, the line shape is therefore controlled by the boundary-RG coordinate $T_B/T$, where $T_B$ is the boundary crossover scale. The tip height tunes this coordinate through the bare local tunneling parameter,
\begin{equation}
T_B=T_{B,0}\left|
\frac{\lambda_0}{\lambda_{0,\rm ref}}
\right|^{(1+r)/r}.
\end{equation}
As discussed above and derived in App.~\ref{app_TBA}, near the full-transmission fixed point the bare backscattering amplitude is proportional to the reflection amplitude, $\lambda_0\propto \left|(\Gamma_t-\Gamma_e)/(\Gamma_t+\Gamma_e)\right|$. Here $\Gamma_t\propto t_0^2$ is the broadening of the probed state induced by the tip, while $\Gamma_e$ is its escape broadening into the electronic continuum of the sample, such as bulk or substrate states. The dependence on $\Gamma_t$ changes sign across the hybridization-matching point: $\lambda_0$ decreases with increasing $\Gamma_t$ for $\Gamma_t<\Gamma_e$, vanishes at $\Gamma_t=\Gamma_e$, and increases for $\Gamma_t>\Gamma_e$. The dominant channel analyzed here is inferred to lie in the latter regime. Lowering the tip therefore increases $\lambda_0$, raises $T_B/T$, and drives the junction toward the low-energy backscattering-dominated regime.

In STM, this universal DCB response dresses the projected local spectral density $\rho_{\rm tip}(E)$. In the low-temperature, slowly varying-DOS limit used here, the measured conductance is
\begin{equation}
G(V)\simeq \rho_{\rm tip}(eV)
G_{\rm env}(V;T_B,r,T),
\label{dressed_conductance}
\end{equation}
where $G_{\rm env}$ denotes the universal TBA DCB response of a constant-DOS junction. Thus, the tip-height series is fitted by varying $T_B/T$ while using the same universal response to dress the local spectral density. The details of TBA theory are given in App.~\ref{app_TBA}.

\section{Experimental platform and normal-state identification of the Ohmic DCB environment}
Measurements were performed by in situ low-temperature STM/STS on Molecular-beam-epitaxy-grown SrSn$_3$ thin films on Si(111). Unless stated otherwise, all setpoint-dependent spectra discussed below were acquired at $T=400$ mK on the same $37.6$-nm island. For each setpoint series, the spectra were acquired at the same lateral position while the setpoint current was varied at a fixed stabilization bias to tune the tip height. In the weak-tunneling limit and at fixed $V_{\rm set}$, $I_{\rm set}\propto |t_0|^2$ to leading order (see details in App.~\ref{app_expe_control}).

We first establish the dissipative response of the junction in the normal state. At $B=1$ T, superconductivity is suppressed and the wide-bias spectra remain metallic and smooth over the full setpoint range [Fig.~\ref{fig:setup_dcb}(b)]. To be noted, in the normalized low-bias window, however, a reproducible zero-bias valley becomes progressively deeper as the tip is lowered. Its persistence after superconductivity is suppressed identifies the anomaly as the local DCB response, with a fitted dissipation strength $r_{\rm HF}\approx 0.22$ (see details in App.~\ref{app_fit_parameter_summary}), thereby establishing the Ohmic environment used below.

Reference spectra, the exponential tip-height dependence of the junction conductance, and pre/post vortex maps exclude tip instability, contact formation, and vortex motion; the complete controls are given in App.~\ref{app_expe_control}.

\section{Vortex spectra}
Fig.~\ref{fig:sc_vortex_single_channel} shows the central boundary-RG test of the experiment. We first analyze the vortex-center spectra [Fig.~\ref{fig:sc_vortex_single_channel}(a)], acquired at $B=0.1$ T on the same $37.6$-nm island. In the high-tip, weak-coupling regime, the vortex center shows a clean, non-split ZBP on top of the small-gap background (see App.~\ref{app_high_tip_line_scan}), a line shape that would be a Majorana candidate in conventional static STM spectroscopy. A
tip-height flow can serve as a dynamical test. For $r<1/2$, Eq.~\eqref{eq:mzm_conductance} predicts that an isolated MZM flows toward the protected strong-coupling fixed point as the boundary coordinate $T_B/T$ is increased; equivalently, the zero-bias response should not be driven into the ordinary DCB valley. Experimentally, lowering the tip increases $\lambda_0$ and hence $T_B/T$, but the ZBP is continuously suppressed, and finally turns into a zero-bias dip. This flow has the opposite sign to the Majorana scaling in Eq.~\eqref{eq:mzm_conductance} and is the expected DCB response of an ordinary vortex-core resonance or local DOS dressed by $G_{\rm env}$. We therefore identify the observed ZBP in this sample as a non-topological vortex-core feature rather than an isolated MZM.

To quantify this conclusion, we fit the vortex spectra with a minimal topologically trivial spectral ansatz: a broad vortex-core continuum plus a smooth background, both dressed by the universal TBA response $G_{\rm env}$. The single-channel fit captures the central suppression and the peak-to-dip crossover [Figs.~\ref{fig:sc_vortex_single_channel}(b,c)], yielding $r^{s}_{\rm v}\simeq0.19<1/2$. Here we have defined $r^{s(d)}_{\rm v}$ as the single-channel (double-channel) fitted dissipation strength.
This independently places the junction in the Majorana-filter regime and quantitatively supports the trivial boundary flow identified above.
To be noted, the purpose of this fit is to extract the low-energy boundary flow, not to reproduce all finite-bias structures.

\begin{figure}[t]
    \centering
    \includegraphics[width=\linewidth]{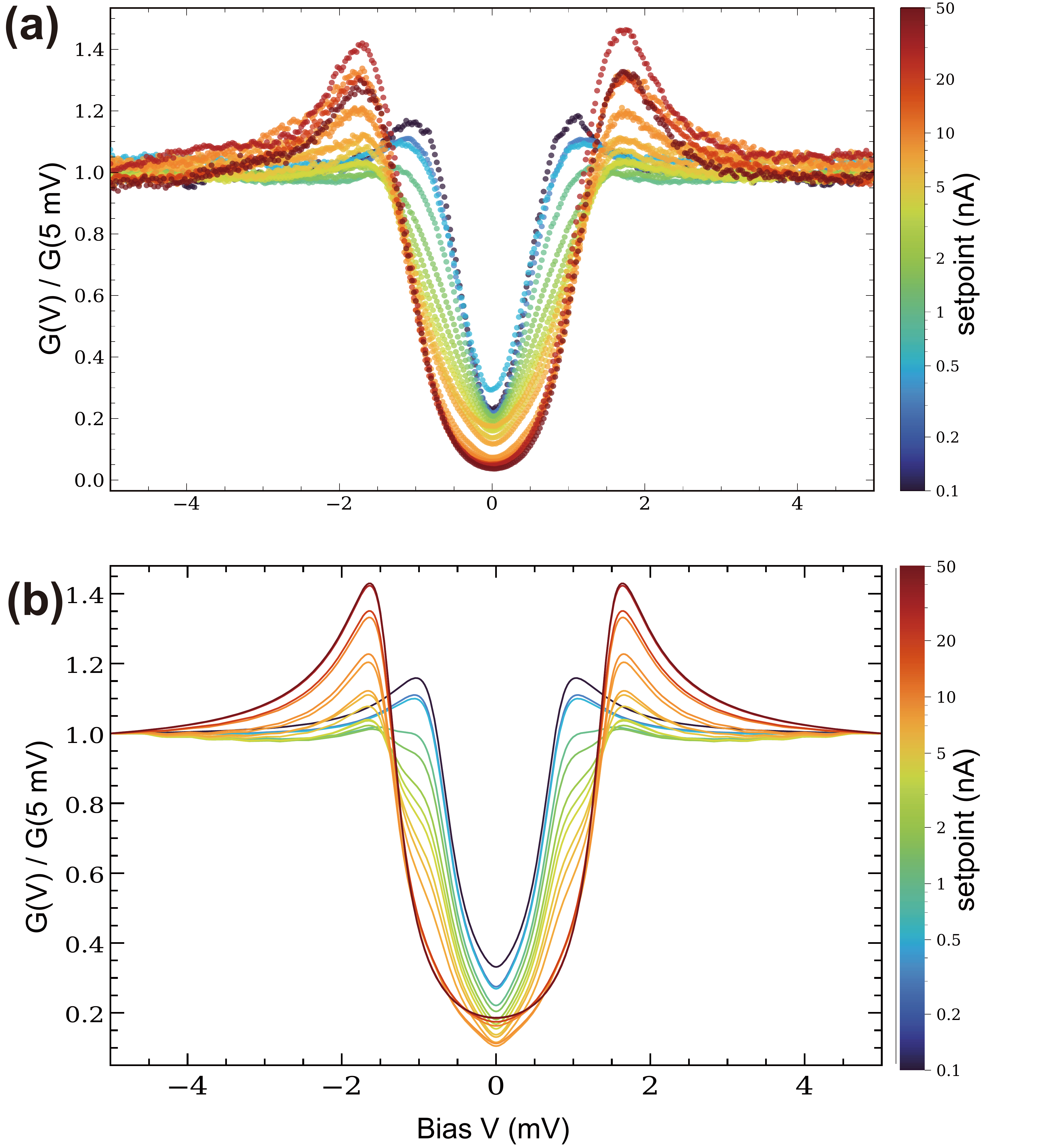}
    \caption{
    \textbf{Superconducting-gap spectra data and the two-channel interpretation.}
    \textbf{(a)} Experimental data of $37.6$ nm superconducting-gap spectra.
    \textbf{(b)} Two-gap fit of the superconducting spectra with Eq.~\eqref{eq:two_gap_sc}, corresponding to a strongly
    tip-coupled small-gap channel and a weakly tip-coupled large-gap channel. The fitted dissipation strength is $r_{\rm SC}\approx 0.21$.
    }
    \label{fig:two_channel}
\end{figure}

The residual discrepancy between the single-continuum fit and the full vortex spectra is informative. A real vortex core can host multiple low-lying CdGM states and additional spectral weight. Because the single-channel analysis targets the low-energy boundary flow, the shoulder-like residual requires another explanation. A parallel analysis of the superconducting-gap spectra shows that DCB suppresses the small-gap coherence peaks in the lower-tip-height regime, consistent with the experimental data [Fig.~\ref{fig:two_channel}(a). However, a trivial single-channel model cannot explain the appearance of the second coherent-peak structure at larger bias. These two residual structures therefore motivate the multichannel DCB contrast analysis below.

\begin{figure}[t]
    \centering
    \includegraphics[width=\linewidth]{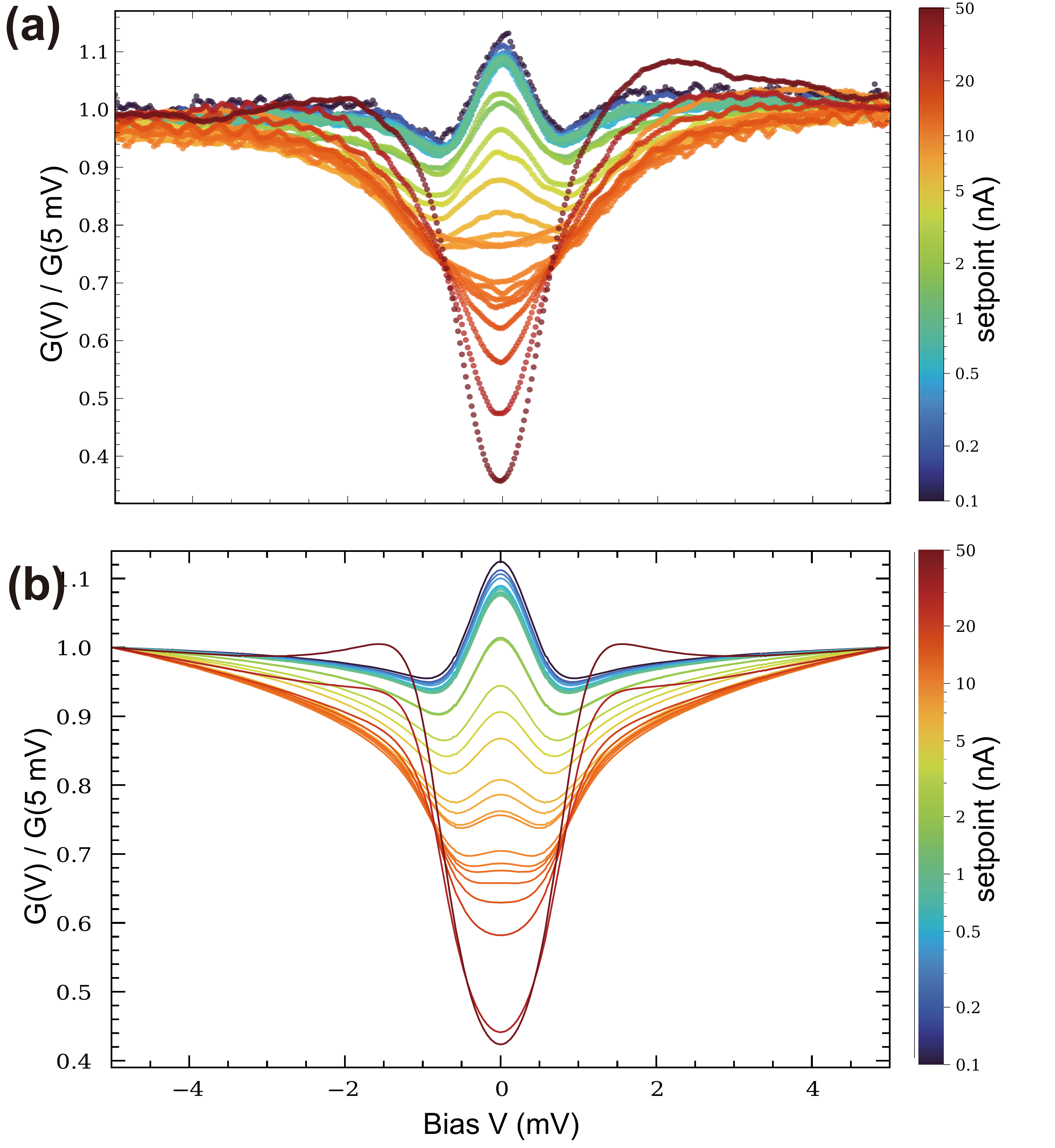}
    \caption{
    \textbf{Two-channel interpretation of the residual spectral weight in vortex spectra.} \textbf{(a)} Experimental data of $37.6$ nm vortex spectra.
    \textbf{(b)} Two-gap fit of the vortex spectra with Eq.~\eqref{eq:vortex_projection}, with an additional residual superconducting channel to the DCB-dressed vortex continuum. The fitted dissipation strength is $r_{\rm v}^d\approx 0.18$.
    }
    \label{fig:vortex_two_channel}
\end{figure}

\section{Multichannel DCB contrast in the gap spectra}
The single-channel analysis above establishes the low-energy boundary flow, but it does not explain every feature in the full spectra. Interestingly, the remaining
inconsistency has a clear structure and has the same structure in two independent data sets. In the vortex-center spectra, the single-continuum fit captures the suppression of the ZBP but underestimates the shoulder-like spectral weight near $|V|\sim 1$--$2~\mathrm{mV}$, which becomes more visible as the tip is lowered, as shown in Figs.~\ref{fig:sc_vortex_single_channel}(a). In the full superconducting spectra, the coherence peaks of the small-gap channel are first suppressed by DCB, while a larger $U$-shaped gap-like contribution emerges at stronger coupling [Figs.~\ref{fig:two_channel}(a)]. This systematic residual motivates a multichannel description and connects naturally to the reported two-component superconductivity of SrSn$_3$~\cite{song2025srsn3,yang2025srsn3}.

We interpret both effects through multichannel DCB contrast. Each local channel $a$ has its own bare backscattering
amplitude $\lambda_a$ and boundary scale, $T_{B,a}\propto |\lambda_a|^{(1+r)/r}$. Different channels can lie on opposite sides of their respective hybridization-matching points. The dominant channel discussed above is overcoupled, $\Gamma_{t,1}>\Gamma_{e,1}$, so lowering the tip increases $\lambda_1$ and drives this channel deeper into the DCB-suppressed regime. A more weakly accessed channel can instead be undercoupled, $\Gamma_{t,2}<\Gamma_{e,2}$, in which case lowering the tip reduces $\lambda_2$ and moves the channel toward the full-transmission point. Its low-energy contribution therefore becomes progressively more visible.

For the superconducting-gap spectra, we fit two projected superconducting components,
\begin{align}
    G_{\mathrm{SC}}(V)=G_1(V;\lambda_1,\Delta_1,\Gamma_1)
    +
    G_2\!\left(V;\lambda_2,\Delta_2,\Gamma_2\right),
    \label{eq:two_gap_sc}
\end{align}
where each channel is dressed by the same environmental response but has its own boundary scale. We model the superconducting gap as Dynes density of states~\cite{dynes1978lifetime,sun2015dirac,du2016scrutinizing} dressed by the universal TBA DCB response, as detailed in App.~\ref{app_smooth_approx}. The fit gives $r_{\rm SC}\simeq0.21$, $\Delta_1\simeq0.76~\mathrm{meV}$ and $\Delta_2\simeq1.44~\mathrm{meV}$.
The agreement between $r_{\mathrm SC}$ and $r^s_{\mathrm v}$ is the key internal consistency check: the superconducting gap renormalization and the vortex-center peak-to-dip evolution are governed by the same Ohmic environment in the same sample.

In the two-channel fit, we impose that $\lambda_1$ increases while $\lambda_2$ decreases monotonically with increasing setpoint current. Over the experimentally accessed range, these trends correspond to an overcoupled first channel, $\Gamma_{t,1}>\Gamma_{e,1}$, and an undercoupled second channel, $\Gamma_{t,2}<\Gamma_{e,2}$. A natural microscopic assignment is that the small-gap component is associated with a near-surface channel that is more strongly accessed by the tip, whereas the larger-gap component originates from a more weakly accessed bulk superconducting channel.

\section{Hidden superconducting spectral weight in the vortex}
The same contrast mechanism accounts for the vortex-center residuals [Fig.~\ref{fig:vortex_two_channel}]. Instead of treating the vortex spectrum as a single DOS, we write the measured conductance as a sum of separately dressed channels,
\begin{equation}
\rho_{\rm tip}^{\rm vortex}(E)
=
\rho_{\rm CdGM}(E)
+
\bar\rho_{\rm SC}(E).
\label{eq:vortex_projection}
\end{equation}
Here the first term describes the strongly accessed vortex-core continuum, while the second term represents weakly projected residual superconducting spectral weight.
As the vortex-core continuum is suppressed by the dissipative flow, the weak superconducting contribution gains relative contrast and appears as the lifted side shoulder. The fitted superconducting scale of this additional vortex component, $\Delta_{\mathrm{vortex,SC}}\simeq0.91~\mathrm{meV}$,
lying within the range of the superconducting gaps fitted above.
This consistency supports interpreting the emerging shoulder as residual superconducting spectral weight.
The fitted dissipation strength, $r_v^{d}\simeq0.18$, further confirms that the different fits probe the same local Ohmic environment.

\section{Discussion}
Our observation demonstrates that an ordinary vortex-core state can generate a clean, non-split ZBP and thereby be misinterpreted as a Majorana signature in static spectroscopy. The present protocol is therefore designed as a falsification test for Majorana-like vortex ZBPs, rather than as a complete classification of vortex topology. The observed trivial boundary flow shows that the dominant spectral weight forming the ZBP is incompatible with the protected response of an isolated MZM.
Applying this protocol to platforms with independently established topological vortex modes and tunable local impedance would enable a direct comparison between protected Majorana fixed-point flow and conventional DCB response. The same dissipative mechanism also provides a possible route to enhancing the relative visibility of weakly coupled superconducting channels.

\begin{acknowledgments}
    \textbf{Acknowledgments}: Z.Z. and D.E.L. are supported by the Quantum Science and Technology-National Science and Technology Major Project (Grant No. 2021ZD0302400). Q.Z., Y.W.W. and C.L.S. are supported by the National Key Research and Development Program of China (Grant No. 2022YFA1403100) and the Natural Science Foundation of China (Grant No. 12474130, Grant No. 12134008, and Grant No. 52388201).
\end{acknowledgments}

\bibliographystyle{apsrev4-2}
\bibliography{ref}

@article{binnig1982controllable_gap,
  title={Tunneling through a controllable vacuum gap},
  author={Binnig, Gerd and Rohrer, Heinrich and Gerber, Ch and Weibel, Eddie},
  journal={Applied Physics Letters},
  volume={40},
  number={2},
  pages={178--180},
  year={1982},
  publisher={AIP Publishing},
  url = {https://pubs.aip.org/aip/apl/article-pdf/40/2/178/18445409/178_1_online.pdf}
}

@article{tersoff1983theory_application,
  title={Theory and application for the scanning tunneling microscope},
  author={Tersoff, Jerry and Hamann, Donald R},
  journal={Physical review letters},
  volume={50},
  number={25},
  pages={1998},
  year={1983},
  publisher={APS},
  url = {https://journals.aps.org/prl/abstract/10.1103/PhysRevLett.50.1998}
}

@article{dynes1978lifetime,
  title={Direct measurement of quasiparticle-lifetime broadening in a strong-coupled superconductor},
  author={Dynes, Robert C and Narayanamurti, Venkatesh and Garno, J Pm},
  journal={Physical Review Letters},
  volume={41},
  number={21},
  pages={1509},
  year={1978},
  publisher={APS},
  url = {https://journals.aps.org/prl/abstract/10.1103/PhysRevLett.41.1509}
}

@article{caroli1964vortex,
  title={Bound fermion states on a vortex line in a type II superconductor},
  author={Caroli, C and De Gennes, P G and Matricon, J},
  journal={Physics Letters},
  volume={9},
  number={4},
  pages={307--309},
  year={1964},
  publisher={Elsevier},
  url = {https://www.sciencedirect.com/science/article/pii/0031916364903750}
}

@article{hess1989vortex,
  title={Scanning-tunneling-microscope observation of the Abrikosov flux lattice and the density of states near and inside a fluxoid},
  author={Hess, H F and Robinson, R B and Dynes, R C and Valles Jr, J M and Waszczak, J V},
  journal={Physical review letters},
  volume={62},
  number={2},
  pages={214},
  year={1989},
  publisher={APS},
  url = {https://journals.aps.org/prl/abstract/10.1103/PhysRevLett.62.214}
}

@article{pan2000vortex,
  title={STM studies of the electronic structure of vortex cores in Bi 2 Sr 2 CaCu 2 O 8+ $\delta$},
  author={Pan, SH and Hudson, EW and Gupta, AK and Ng, K-W and Eisaki, H and Uchida, S and Davis, JC},
  journal={Physical Review Letters},
  volume={85},
  number={7},
  pages={1536},
  year={2000},
  publisher={APS},
  url = {https://journals.aps.org/prl/abstract/10.1103/PhysRevLett.85.1536}
}

@article{kitaev2001unpaired,
  title={Unpaired Majorana fermions in quantum wires},
  author={Kitaev, A Yu},
  journal={Physics-uspekhi},
  volume={44},
  number={10S},
  pages={131--136},
  year={2001},
  url = {https://iopscience.iop.org/article/10.1070/1063-7869/44/10S/S29/meta}
}

@article{readgreen2000paired,
  title={Paired states of fermions in two dimensions with breaking of parity and time-reversal symmetries and the fractional quantum Hall effect},
  author={Read, Nicholas and Green, Dmitry},
  journal={Physical Review B},
  volume={61},
  number={15},
  pages={10267},
  year={2000},
  publisher={APS},
  url = {https://journals.aps.org/prb/abstract/10.1103/PhysRevB.61.10267}
}

@article{fu2008proximity,
  title={Superconducting proximity effect and majorana fermions at the surface of a topological insulator},
  author={Fu, Liang and Kane, Charles L},
  journal={Physical review letters},
  volume={100},
  number={9},
  pages={096407},
  year={2008},
  publisher={APS},
  url = {https://journals.aps.org/prl/abstract/10.1103/PhysRevLett.100.096407}
}

@article{alicea2012directions,
  title={New directions in the pursuit of Majorana fermions in solid state systems},
  author={Alicea, Jason},
  journal={Reports on progress in physics},
  volume={75},
  number={7},
  pages={076501},
  year={2012},
  publisher={IOP Publishing},
  url = {https://iopscience.iop.org/article/10.1088/0034-4885/75/7/076501/meta}
}

@article{wang2018fetse_majorana,
  title={Evidence for Majorana bound states in an iron-based superconductor},
  author={Wang, Dongfei and Kong, Lingyuan and Fan, Peng and Chen, Hui and Zhu, Shiyu and Liu, Wenyao and Cao, Lu and Sun, Yujie and Du, Shixuan and Schneeloch, John and others},
  journal={Science},
  volume={362},
  number={6412},
  pages={333--335},
  year={2018},
  publisher={American Association for the Advancement of Science},
  url = {https://www.science.org/doi/abs/10.1126/science.aao1797}
}

@article{machida2019zero_energy,
  title={Zero-energy vortex bound state in the superconducting topological surface state of Fe (Se, Te)},
  author={Machida, T and Sun, Y and Pyon, S and Takeda, S and Kohsaka, Y and Hanaguri, T and Sasagawa, T and Tamegai, T},
  journal={Nature materials},
  volume={18},
  number={8},
  pages={811--815},
  year={2019},
  publisher={Nature Publishing Group UK London},
  url = {https://www.nature.com/articles/s41563-019-0397-1}
}

@article{zhu2020quantized,
  title={Nearly quantized conductance plateau of vortex zero mode in an iron-based superconductor},
  author={Zhu, Shiyu and Kong, Lingyuan and Cao, Lu and Chen, Hui and Papaj, Micha{\l} and Du, Shixuan and Xing, Yuqing and Liu, Wenyao and Wang, Dongfei and Shen, Chengmin and others},
  journal={Science},
  volume={367},
  number={6474},
  pages={189--192},
  year={2020},
  publisher={American Association for the Advancement of Science},
  url = {https://www.science.org/doi/abs/10.1126/science.aax0274}
}

@article{liu2012zbp,
  title={Zero-Bias Peaks in the Tunneling Conductance of Spin-Orbit-Coupled Superconducting Wires with and without Majorana End-States},
  author={Liu, Jie and Potter, Andrew C and Law, KT and Lee, Patrick A},
  journal={Physical review letters},
  volume={109},
  number={26},
  pages={267002},
  year={2012},
  publisher={APS},
  url = {https://journals.aps.org/prl/abstract/10.1103/PhysRevLett.109.267002}
}

@article{kells2012trivial,
  title={Near-zero-energy end states in topologically trivial spin-orbit coupled superconducting nanowires with a smooth confinement},
  author={Kells, Graham and Meidan, Dganit and Brouwer, Piet W},
  journal={Physical Review B—Condensed Matter and Materials Physics},
  volume={86},
  number={10},
  pages={100503(R)},
  year={2012},
  publisher={APS},
  url = {https://journals.aps.org/prb/abstract/10.1103/PhysRevB.86.100503}
}

@article{prada2012transport,
  title={Transport spectroscopy of NS nanowire junctions with Majorana fermions},
  author={Prada, Elsa and San-Jose, Pablo and Aguado, Ram{\'o}n},
  journal={Physical Review B—Condensed Matter and Materials Physics},
  volume={86},
  number={18},
  pages={180503(R)},
  year={2012},
  publisher={APS},
  url = {https://journals.aps.org/prb/abstract/10.1103/PhysRevB.86.180503}
}

@article{caldeira1983quantum,
  title={Quantum tunnelling in a dissipative system},
  author={Caldeira, Amir O and Leggett, Anthony J},
  journal={Annals of physics},
  volume={149},
  number={2},
  pages={374--456},
  year={1983},
  publisher={Academic Press},
  url = {https://www.sciencedirect.com/science/article/abs/pii/0003491683902026}
}

@article{schmid1983diffusion,
  title={Diffusion and localization in a dissipative quantum system},
  author={Schmid, Albert},
  journal={Physical Review Letters},
  volume={51},
  number={17},
  pages={1506},
  year={1983},
  publisher={APS},
  url = {https://journals.aps.org/prl/abstract/10.1103/PhysRevLett.51.1506}
}

@article{devoret1990electromagnetic,
  title={Effect of the electromagnetic environment on the Coulomb blockade in ultrasmall tunnel junctions},
  author={ Devoret, M. H.  and  Esteve, D.  and  Grabert, H.  and  Ingold, G. L.  and  Pothier, H.  and  Urbina, C. },
  journal={Physical Review Letters},
  volume={64},
  number={15},
  pages={1824},
  year={1990},
  url = {https://journals.aps.org/prl/abstract/10.1103/PhysRevLett.64.1824}
}

@article{girvin1990quantum,
  title={Quantum fluctuations and the single-junction Coulomb blockade},
  author={ Girvin, S. M.  and  Glazman, L. I.  and  Jonson, M.  and  Penn, D. R.  and  Stiles, M. D. },
  journal={Physical Review Letters},
  volume={64},
  number={26},
  pages={3183-3186},
  year={1990},
  url = {https://journals.aps.org/prl/abstract/10.1103/PhysRevLett.64.3183}
}

@incollection{ingold1992charge,
  title={Charge tunneling rates in ultrasmall junctions},
  author={Ingold, Gert-Ludwig and Nazarov, Yu V},
  booktitle={Single charge tunneling: Coulomb blockade phenomena in nanostructures},
  pages={21--107},
  year={1992},
  publisher={Springer},
  url = {https://link.springer.com/chapter/10.1007/978-1-4757-2166-9_2}
}

@article{kane1992barriers,
  title={Transmission through barriers and resonant tunneling in an interacting one-dimensional electron gas},
  author={Kane, CL and Fisher, Matthew PA},
  journal={Physical Review B},
  volume={46},
  number={23},
  pages={15233},
  year={1992},
  publisher={APS},
  url = {https://journals.aps.org/prb/abstract/10.1103/PhysRevB.46.15233}
}

@article{fendley1995exact,
  title={Exact nonequilibrium transport through point contacts in quantum wires and fractional quantum Hall devices},
  author={Fendley, P and Ludwig, AWW and Saleur, H},
  journal={Physical Review B},
  volume={52},
  number={12},
  pages={8934},
  year={1995},
  publisher={APS},
  url = {https://journals.aps.org/prb/abstract/10.1103/PhysRevB.52.8934}
}

@article{safi2004one_channel,
  title={One-channel conductor in an ohmic environment: mapping to a Tomonaga-Luttinger liquid and full counting statistics},
  author={Safi, Ines and Saleur, Hubert},
  journal={Physical review letters},
  volume={93},
  number={12},
  pages={126602},
  year={2004},
  publisher={APS},
  url = {https://journals.aps.org/prl/abstract/10.1103/PhysRevLett.93.126602}
}

@article{parmentier2011backaction,
  title={Strong back-action of a linear circuit on a single electronic quantum channel},
  author={Parmentier, FD and Anthore, A and Jezouin, S and Le Sueur, H and Gennser, U and Cavanna, Antonella and Mailly, D and Pierre, F},
  journal={Nature Physics},
  volume={7},
  number={12},
  pages={935--938},
  year={2011},
  publisher={Nature Publishing Group UK London},
  url = {https://www.nature.com/articles/nphys2092}
}

@article{brun2012dcb_contacts,
  title={Dynamical Coulomb blockade observed in nanosized electrical contacts},
  author={Brun, Christophe and M{\"u}ller, Konrad H and Hong, I-Po and Patthey, Fran{\c{c}}ois and Flindt, Christian and Schneider, Wolf-Dieter},
  journal={Physical review letters},
  volume={108},
  number={12},
  pages={126802},
  year={2012},
  publisher={APS},
  url = {https://journals.aps.org/prl/abstract/10.1103/PhysRevLett.108.126802}
}

@article{mebrahtu2013majorana_critical,
  title={Observation of Majorana quantum critical behaviour in a resonant level coupled to a dissipative environment},
  author={Mebrahtu, HT and Borzenets, IV and Zheng, H and Bomze, Yu V and Smirnov, AI and Florens, Serge and Baranger, HU and Finkelstein, G},
  journal={Nature Physics},
  volume={9},
  number={11},
  pages={732--737},
  year={2013},
  publisher={Nature Publishing Group UK London},
  url = {https://www.nature.com/articles/nphys2735}
}

@article{senkpiel2020local_probe,
  title={Dynamical Coulomb blockade as a local probe for quantum transport},
  author={Senkpiel, Jacob and Kl{\"o}ckner, Jan C and Etzkorn, Markus and Dambach, Simon and Kubala, Bj{\"o}rn and Belzig, Wolfgang and Yeyati, Alfredo Levy and Cuevas, Juan Carlos and Pauly, Fabian and Ankerhold, Joachim and others},
  journal={Physical Review Letters},
  volume={124},
  number={15},
  pages={156803},
  year={2020},
  publisher={APS},
  url = {https://journals.aps.org/prl/abstract/10.1103/PhysRevLett.124.156803}
}

@article{liu2013majorana_filter,
  title={Proposed Method for Tunneling Spectroscopy with Ohmic Dissipation Using Resistive Electrodes: A Possible Majorana Filter},
  author={Liu, Dong E},
  journal={Physical Review Letters},
  volume={111},
  number={20},
  pages={207003},
  year={2013},
  publisher={APS},
  url = {https://journals.aps.org/prl/abstract/10.1103/PhysRevLett.111.207003}
}

@article{liu2022universal,
  title={Universal conductance scaling of Andreev reflections using a dissipative probe},
  author={Liu, Donghao and Zhang, Gu and Cao, Zhan and Zhang, Hao and Liu, Dong E},
  journal={Physical Review Letters},
  volume={128},
  number={7},
  pages={076802},
  year={2022},
  publisher={APS},
  url = {https://journals.aps.org/prl/abstract/10.1103/PhysRevLett.128.076802}
}

@article{zhang2023vortex_dissipative,
  title={Theoretical proposal to obtain strong Majorana evidence from scanning tunneling spectroscopy of a vortex with a dissipative environment},
  author={Zhang, Gu and Li, Chuang and Li, Geng and Song, Can-Li and Liu, Xin and Zhang, Fu-Chun and Liu, Dong E},
  journal={Physical Review B},
  volume={107},
  number={19},
  pages={195413},
  year={2023},
  publisher={APS},
  url = {https://journals.aps.org/prb/abstract/10.1103/PhysRevB.107.195413}
}

@article{zhang2021conductance,
  title={Conductance of a dissipative quantum dot: Nonequilibrium crossover near a non-Fermi-liquid quantum critical point},
  author={Zhang, Gu and Novais, Eduardo and Baranger, Harold U},
  journal={Physical Review B},
  volume={104},
  number={16},
  pages={165423},
  year={2021},
  publisher={APS},
  url={https://journals.aps.org/prb/abstract/10.1103/PhysRevB.104.165423}
}

@article{zhang2022dissipative_lead,
  title={Suppressing Andreev bound state zero bias peaks using a strongly dissipative lead},
  author={Zhang, Shan and Wang, Zhichuan and Pan, Dong and Li, Hangzhe and Lu, Shuai and Li, Zonglin and Zhang, Gu and Liu, Donghao and Cao, Zhan and Liu, Lei and others},
  journal={Physical Review Letters},
  volume={128},
  number={7},
  pages={076803},
  year={2022},
  publisher={APS},
  url = {https://journals.aps.org/prl/abstract/10.1103/PhysRevLett.128.076803}
}

@article{wang2022weak_dissipation,
  title={Large Andreev bound state zero-bias peaks in a weakly dissipative environment},
  author={Wang, Zhichuan and Zhang, Shan and Pan, Dong and Zhang, Gu and Xia, Zezhou and Li, Zonglin and Liu, Donghao and Cao, Zhan and Liu, Lei and Wen, Lianjun and others},
  journal={Physical Review B},
  volume={106},
  number={20},
  pages={205421},
  year={2022},
  publisher={APS},
  url = {https://journals.aps.org/prb/abstract/10.1103/PhysRevB.106.205421}
}

@article{zhang2023insitu_dcb,
  title={In situ tuning of dynamical Coulomb blockade on Andreev bound states in hybrid nanowire devices},
  author={Zhang, Shan and Wang, Zhichuan and Pan, Dong and Wang, Zhaoyu and Li, Zonglin and Zhang, Zitong and Gao, Yichun and Cao, Zhan and Zhang, Gu and Liu, Lei and others},
  journal={Physical Review B},
  volume={108},
  number={23},
  pages={235416},
  year={2023},
  publisher={APS},
  url = {https://journals.aps.org/prb/abstract/10.1103/PhysRevB.108.235416}
}

@article{song2025srsn3,
  title={Surface superconductivity and topological band in the strong-coupling superconductor SrSn 3},
  author={Song, Jiangpeng and Lan, Xiaobin and Han, Yuyan and Xi, Chuanying and Zhang, Lei and Wu, Peng and Cao, Liang and Liu, Dayong and Xiong, Yimin},
  journal={Physical Review B},
  volume={112},
  number={9},
  pages={094508},
  year={2025},
  publisher={APS},
  url = {https://link.aps.org/doi/10.1103/wttr-2ym9}
}

@article{yang2025srsn3,
  title={Anisotropic superconductivity in the topological semimetal SrSn 3 with layered Sn kagome network},
  author={Yang, Yang and Li, Jiayang and Rehman, Majeed Ur and Ning, Wei and Hao, Ning and Zhu, Xiangde and Tian, Mingliang},
  journal={Physical Review B},
  volume={111},
  number={18},
  pages={184517},
  year={2025},
  publisher={APS},
  url = {https://journals.aps.org/prb/abstract/10.1103/PhysRevB.111.184517}
}

@article{pan2020physical_zbp,
  title={Physical mechanisms for zero-bias conductance peaks in Majorana nanowires},
  author={Pan, Haining and Das Sarma, S},
  journal={Physical Review Research},
  volume={2},
  number={1},
  pages={013377},
  year={2020},
  publisher={APS},
  url = {https://journals.aps.org/prresearch/abstract/10.1103/PhysRevResearch.2.013377}
}

@article{yu2021nonmajorana,
  title={Non-Majorana states yield nearly quantized conductance in proximatized nanowires},
  author={Yu, Peng and Chen, J and Gomanko, M and Badawy, G and Bakkers, EPAM and Zuo, K and Mourik, V and Frolov, SM},
  journal={Nature Physics},
  volume={17},
  number={4},
  pages={482--488},
  year={2021},
  publisher={Nature Publishing Group UK London},
  url = {https://www.nature.com/articles/s41567-020-01107-w}
}

@article{pan2021quantized_unquantized,
  title={Quantized and unquantized zero-bias tunneling conductance peaks in Majorana nanowires: Conductance below and above 2 e 2/h},
  author={Pan, Haining and Liu, Chun-Xiao and Wimmer, Michael and Das Sarma, Sankar},
  journal={Physical Review B},
  volume={103},
  number={21},
  pages={214502},
  year={2021},
  publisher={APS},
  url = {https://journals.aps.org/prb/abstract/10.1103/PhysRevB.103.214502}
}

@article{valentini2021nontopological,
  title={Nontopological zero-bias peaks in full-shell nanowires induced by flux-tunable Andreev states},
  author={Valentini, Marco and Penaranda, Fernando and Hofmann, Andrea and Brauns, Matthias and Hauschild, Robert and Krogstrup, Peter and San-Jose, Pablo and Prada, Elsa and Aguado, Ram{\'o}n and Katsaros, Georgios},
  journal={Science},
  volume={373},
  number={6550},
  pages={82--88},
  year={2021},
  publisher={American Association for the Advancement of Science},
  url = {https://www.science.org/doi/abs/10.1126/science.abf1513}
}

@article{song2026charge_stripes_mzm,
  title={Intertwined Charge Stripes and Majorana Zero Modes in An Iron-Based Superconductor},
  author={Liu, Yu and Wei, Li-Xuan and Cheng, Qiang-Jun and Zhu, Zhenhua and Shi, Xin-Yu and Lou, Cong-Cong and Wang, Yong-Wei and Deng, Ze-Xian and Ren, Ming-Qiang and Liu, Dong E and others},
  journal={arXiv preprint arXiv:2601.15873},
  year={2026},
  url = {https://arxiv.org/abs/2601.15873}
}

@article{serrier2013scanning,
  title={Scanning Tunneling Spectroscopy Study of the Proximity Effect in a Disordered Two-Dimensional Metal},
  author={Serrier-Garcia, L and Cuevas, JC and Cren, Ten and Brun, Christophe and Cherkez, V and Debontridder, Fran{\c{c}}ois and Fokin, D and Bergeret, FS and Roditchev, D},
  journal={Physical review letters},
  volume={110},
  number={15},
  pages={157003},
  year={2013},
  publisher={APS},
  url={https://journals.aps.org/prl/abstract/10.1103/PhysRevLett.110.157003}
}

@article{sun2015dirac,
  title={Dirac surface states and nature of superconductivity in noncentrosymmetric BiPd},
  author={Sun, Zhixiang and Enayat, Mostafa and Maldonado, Ana and Lithgow, Calum and Yelland, Ed and Peets, Darren C and Yaresko, Alexander and Schnyder, Andreas P and Wahl, Peter},
  journal={Nature Communications},
  volume={6},
  number={1},
  pages={6633},
  year={2015},
  publisher={Nature Publishing Group UK London},
  url={https://www.nature.com/articles/ncomms7633}
}

@article{du2016scrutinizing,
  title={Scrutinizing the double superconducting gaps and strong coupling pairing in (Li1- x Fe x) OHFeSe},
  author={Du, Zengyi and Yang, Xiong and Lin, Hai and Fang, Delong and Du, Guan and Xing, Jie and Yang, Huan and Zhu, Xiyu and Wen, Hai-Hu},
  journal={Nature communications},
  volume={7},
  number={1},
  pages={10565},
  year={2016},
  publisher={Nature Publishing Group UK London},
  url={https://www.nature.com/articles/ncomms10565}
}

@article{liu2018robust,
  title={Robust and clean Majorana zero mode in the vortex core of high-temperature superconductor (Li 0.84 Fe 0.16) OHFeSe},
  author={Liu, Qin and Chen, Chen and Zhang, Tong and Peng, Rui and Yan, Ya-Jun and Wen, Chen-Hao-Ping and Lou, Xia and Huang, Yu-Long and Tian, Jin-Peng and Dong, Xiao-Li and others},
  journal={Physical Review X},
  volume={8},
  number={4},
  pages={041056},
  year={2018},
  publisher={APS},
  url={https://journals.aps.org/prx/abstract/10.1103/PhysRevX.8.041056}
}

@article{kong2019half,
  title={Half-integer level shift of vortex bound states in an iron-based superconductor},
  author={Kong, Lingyuan and Zhu, Shiyu and Papaj, Micha{\l} and Chen, Hui and Cao, Lu and Isobe, Hiroki and Xing, Yuqing and Liu, Wenyao and Wang, Dongfei and Fan, Peng and others},
  journal={Nature Physics},
  volume={15},
  number={11},
  pages={1181--1187},
  year={2019},
  publisher={Nature Publishing Group UK London},
  url={https://www.nature.com/articles/s41567-019-0630-5}
}

@article{law2009majorana,
  title={Majorana fermion induced resonant Andreev reflection},
  author={Law, Kam Tuen and Lee, Patrick A and Ng, Tai Kai},
  journal={Physical review letters},
  volume={103},
  number={23},
  pages={237001},
  year={2009},
  publisher={APS},
  url={https://journals.aps.org/prl/abstract/10.1103/PhysRevLett.103.237001}
}

@article{kim2021anisotropic,
  title={Anisotropic non-split zero-energy vortex bound states in a conventional superconductor},
  author={Kim, Howon and Nagai, Yuki and R{\'o}zsa, Levente and Schreyer, Dominik and Wiesendanger, Roland},
  journal={Applied Physics Reviews},
  volume={8},
  number={3},
  year={2021},
  publisher={AIP Publishing},
  url={https://pubs.aip.org/aip/apr/article/8/3/031417/124951}
}

@article{boulat2019full,
  title={Full exact solution of the out-of-equilibrium boundary sine Gordon model},
  author={Boulat, Edouard},
  journal={arXiv preprint arXiv:1912.03872},
  year={2019},
  url={https://arxiv.org/abs/1912.03872}
}

@article{flensberg2010tunneling,
  title={Tunneling characteristics of a chain of Majorana bound states},
  author={Flensberg, Karsten},
  journal={Physical Review B—Condensed Matter and Materials Physics},
  volume={82},
  number={18},
  pages={180516},
  year={2010},
  publisher={APS},
  url={https://journals.aps.org/prb/abstract/10.1103/PhysRevB.82.180516}
}

@article{liu2024dynamical,
  title={Dynamical Coulomb blockade as a signature of the sign-reversing Cooper pairing potential},
  author={Liu, Chaofei and Portugal, Pedro and Gao, Yi and Yang, Jie and Zhang, Xiuying and Liu, Yanzhao and Wei, Tianheng and Ren, Wei and Lu, Jing and Flindt, Christian and others},
  journal={Physical Review B},
  volume={110},
  number={1},
  pages={014514},
  year={2024},
  publisher={APS},
  url={https://journals.aps.org/prb/abstract/10.1103/PhysRevB.110.014514}
}

@article{kawakami2015evolution,
  title={Evolution of density of states and a spin-resolved checkerboard-type pattern associated with the Majorana bound state},
  author={Kawakami, Takuto and Hu, Xiao},
  journal={Physical Review Letters},
  volume={115},
  number={17},
  pages={177001},
  year={2015},
  publisher={APS},
  url={https://journals.aps.org/prl/abstract/10.1103/PhysRevLett.115.177001}
}

\appendix

\section{Experimental control}\label{app_expe_control}
For each setpoint series, the stabilization bias $V_{\rm set}$ was kept fixed while the setpoint current $I_{\rm set}$ was varied stepwise to tune the tip--sample separation. The experimentally used setpoint range was chosen such that the resulting spectral evolution could be resolved while the junction remained in the tunneling regime: throughout this range, the junction current retained its expected exponential dependence on tip height, the junction conductance remained well below the point-contact regime, and no abrupt changes indicative of tip instability or contact formation were observed. At each setpoint, the feedback loop was first used to stabilize the junction at $(V_{\rm set},I_{\rm set})$ and was then disabled during the acquisition of the $dI/dV$ spectrum, so that each spectrum was measured at a fixed tip height. All spectra within a given setpoint series were acquired at the same lateral position.

Atomically resolved topographs confirm a flat SrSn$_3$ surface without an additional reconstruction in the energy window relevant to the present measurements. Pb reference spectra remain unchanged over the setpoint range used here, while spectra acquired on clean SrSn$_3$ terraces are spatially reproducible. The junction conductance follows the expected exponential dependence on tip height and remains well below the point-contact regime. Vortex maps acquired before and after the setpoint-dependent spectroscopy show no detectable displacement of the measured vortex. The spectra shown in this paper are measured inside a 150 nm$\times$150 nm region, while the size of sample is more than 500 nm$\times$500 nm.

\section{High-tip-height vortex line scan: a Majorana-like
static signature}
\label{app_high_tip_line_scan}

To assess how the vortex signal would be interpreted in a conventional STM measurement, we measured a spatially resolved line scan across the vortex at the largest tip--sample separation used in this work. In this high-tip-height regime, the backscattering strength is small and the corresponding boundary scale $T_B$ is close to the strong-tunneling limit. The spectra therefore provide a close experimental approximation to a static measurement of the intrinsic vortex-core local density of states.

\begin{figure*}[t]
    \centering
    \includegraphics[width=0.5\linewidth]{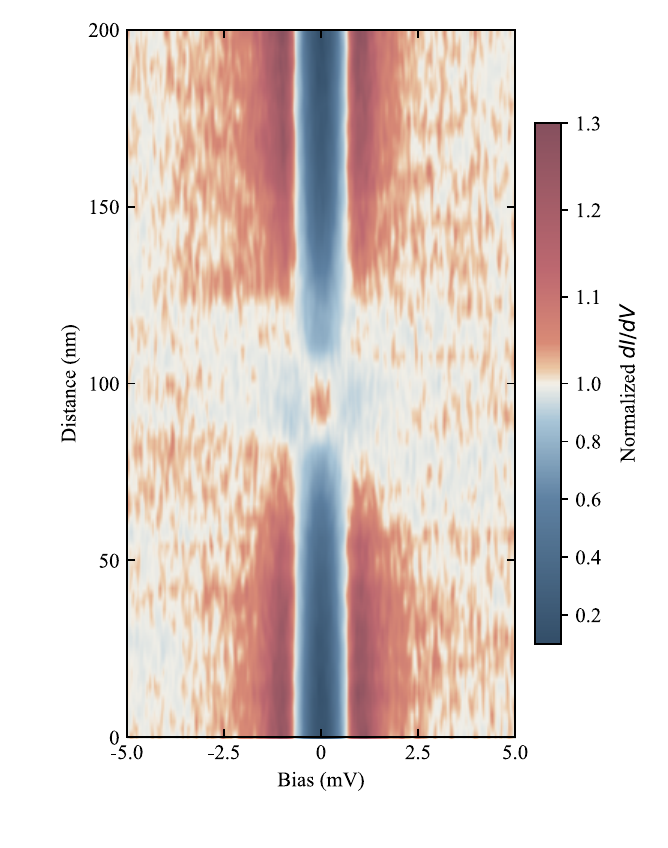}
    \caption{
    High-tip-height selected spectra at representative positions approaching the vortex center (distance = 100). At the vortex center, the spectrum displays a
    localized, non-split zero-bias peak. From a conventional static
    STM perspective, this feature satisfies the usual phenomenological
    criteria of a Majorana-like vortex signal. The subsequent
    tip-height-dependent boundary flow, however, provides the
    dynamical criterion that distinguishes this trivial zero-bias
    resonance from a protected Majorana response.
    }
    \label{fig:high-tip-vortex-line-scan}
\end{figure*}

Fig.~\ref{fig:high-tip-vortex-line-scan} shows the differential
conductance spectra acquired along a line passing through the vortex
center. At the vortex center, the low-energy spectrum exhibits a
single, non-split zero-bias peak on top of the superconducting
background. The peak is spatially localized in the vortex core and
remains centered at zero bias within the experimental energy
resolution. Based on these conventional static criteria alone, the
signal would naturally be regarded as Majorana-like: it is a
localized, non-split vortex-center zero-bias feature without an
apparent finite-energy splitting.

Importantly, such a static line scan cannot by itself establish
topological protection. A trivial vortex-core resonance, including
unresolved Caroli--de Gennes--Matricon states or other low-energy
bound states, can produce a closely similar spatial and spectral
signature in the weak-tunneling regime. The decisive information in
the present experiment instead comes from the controlled
tip-height-dependent boundary flow. As the tip is lowered and
$T_B/T$ is increased, the initially Majorana-like zero-bias peak is
suppressed and ultimately evolves into an ordinary DCB-induced
zero-bias dip. This behavior is incompatible with the protected
strong-coupling flow expected for an isolated Majorana mode in the
$r<1/2$ regime.

\section{Thermodynamic Bethe Ansatz formulation}\label{app_TBA}
We start from the local tunneling Hamiltonian for an STM junction (i.e., the effective lead-dot-lead device) coupled to an Ohmic electromagnetic environment,
\begin{equation}
    H_t =
    t_t \psi^\dagger_{\rm t}(0) e^{-i\varphi/2}d
    + t_e \psi^\dagger_{\rm e}(0)e^{i\varphi/2}d+h.c.,
    \label{eq:end_Ht}
\end{equation}
where $c_{\rm t(e)}$ annihilates a tip (sample electronic reservoir) electron at the tunneling point, $d$ annihilates the sample electron coupled to the tip, and $e^{\mp i\varphi}$ transfers charge to the environmental circuit. The broadening induced by the two channels are $\Gamma_{t(e)}\propto|t_{t(e)}|^2$, respectively. For an Ohmic environment,
\begin{equation}
    \left\langle e^{-i\varphi(t)}e^{i\varphi(0)}\right\rangle
    \propto t^{-2r},\qquad r=R/R_Q,\quad R_Q=h/e^2 .
\end{equation}
We can bosonize the two fermion fields,
\begin{equation}
    \psi_\alpha(x)
=
\frac{F_\alpha}{\sqrt{2\pi a_0}}
e^{i\Phi_\alpha(x)},
\qquad
\alpha=t,e,
\label{eq:lead_bosonization}
\end{equation}
where $F_\alpha$ are Klein factors and $a_0$ is a short-distance cutoff.
By introducing the charge and flavor combinations
\begin{equation}
\Phi_c
=
\frac{\Phi_t+\Phi_e}{\sqrt{2}},
\qquad
\Phi_f
=
\frac{\Phi_t-\Phi_e}{\sqrt{2}},
\label{eq:chiral_charge_flavor}
\end{equation}
the tunneling term becomes
\begin{align}
H_{t}
={}&
\frac{e^{-i\Phi_c(0)/\sqrt{2}}}{\sqrt{2\pi a_0}}
\left\{
t_tF_t^\dagger
e^{-i[\Phi_f(0)/\sqrt{2}+\phi/2]}
\right.
\nonumber\\
&\left.
\hspace{20mm}
+
t_eF_e^\dagger
e^{+i[\Phi_f(0)/\sqrt{2}+\phi/2]}
\right\}d
+
{\rm h.c.}
\label{eq:bosonized_hybridization}
\end{align}

Eq.~\eqref{eq:bosonized_hybridization} describes the weak-link representation near the decoupled-level fixed point.
Following Ref.~\cite{zhang2021conductance}, once the RG flow reaches the vicinity of the full-transmission fixed point, the
local state is strongly hybridized with both reservoirs and is no longer an independent low-energy degree of freedom. The
junction can then be represented as a continuous one-dimensional channel containing two weak backscattering barriers. To connect this strong-coupling representation to the fields introduced above, we identify the tip and sample branches with the right- and left-moving fields of the continuous channel,
\begin{equation}
\psi_{R/L}^\dagger(x)
=
\frac{F^\dagger}{\sqrt{2\pi a_0}}
e^{\pm ik_Fx}
e^{i\sqrt{\pi}[\varphi(x)\pm\theta(x)]}.
\label{eq:nonchiral_bosonization}
\end{equation}
Up to an overall sign convention and constant Klein-factor phases, comparison with Eq.~\eqref{eq:lead_bosonization} gives
\begin{equation}
\theta(x)
=
-\frac{\Phi_f(x)}{\sqrt{2\pi}},
\qquad
\varphi(x)
=
-\frac{\Phi_c(x)}{\sqrt{2\pi}}.
\label{eq:field_correspondence}
\end{equation}
On the half line $x>0$, the fields used in the full-transmission description are therefore
\begin{equation}
\theta_{c/f}(x)
=
\frac{\theta(x)\pm\theta(-x)}{2},
\qquad
\varphi_{c/f}(x)
=
\frac{\varphi(x)\pm\varphi(-x)}{2}.
\label{eq:folded_fields}
\end{equation}

After the quadratic electromagnetic environment is integrated out, the charge-transfer sector is characterized by
\begin{equation}
g=\frac{1}{1+r}.
\end{equation}
We denote the resulting environment-dressed canonical pair by $\theta_c'$ and $\varphi_f'$. The primes indicate the renormalized Gaussian correlations of these fields. The effective Hamiltonian near the full-transmission fixed point is
\begin{align}
H&_{\rm FT}
=\nonumber\\
&\frac{1}{2}\int_0^\infty dx
\left[
(\partial_x\theta_f)^2
+
(\partial_x\varphi_c)^2
+
\frac{1}{g}(\partial_x\theta_c')^2
+
g(\partial_x\varphi_f')^2
\right]
\nonumber\\
&+
A\cos\left[2\sqrt{\pi}\theta_c'(0)\right]
\cos\left[2\sqrt{\pi}\theta_f(\ell/2)+\xi\right]
\nonumber\\
&+
B\sin\left[2\sqrt{\pi}\theta_c'(0)\right]
\sin\left[2\sqrt{\pi}\theta_f(\ell/2)+\xi\right],
\label{eq:full_transmission_hamiltonian}
\end{align}
where $A$ is the symmetric backscattering component, $B$ measures the asymmetry between the two effective barriers, and $\xi$ parameterizes detuning from resonance. At the symmetric resonant point, $B=0$ and $\xi=\pi/2$, and the barrier perturbation is irrelevant. Away from this fine-tuned point, the flavor field $\theta_f$ is frozen below the corresponding crossover scale, so its expectation value can be absorbed into an effective amplitude and phase. Omitting the decoupled free sector then gives the single-barrier model
\begin{equation}
\begin{split}
H_{\rm BSG}
=
\frac{1}{2}\int_0^\infty dx
&\left[
\frac{1}{g}(\partial_x\theta_c')^2
+
g(\partial_x\varphi_f')^2
\right]
\\
&+
\lambda
\cos\left[
2\sqrt{\pi}\theta_c'(0)+\delta
\right].
\end{split}
\label{eq:BSG_theta}
\end{equation}
After shifting the field to absorb $\delta$, rescaling the nonuniversal mode velocity, and defining
\begin{equation}
\Phi(x)
=
2\sqrt{2\pi}\,\theta_c'(x),
\end{equation}
with $\Pi$ the corresponding canonically conjugate momentum, Eq.~\eqref{eq:BSG_theta} can be mapped to a massless boundary sine-Gordon model,
\begin{equation}
\begin{split}
    H_{\rm BSG}
    =
    \frac{1}{8\pi g}\int_0^\infty dx\,
    \left[(\partial_x\Phi)^2+\Pi^2\right]
    +\lambda\cos\!\left[\frac{\Phi(0)}{\sqrt{2}}\right],\\
    g=\frac{1}{1+r}.
\end{split}
\label{eq:end_BSG}
\end{equation}
The tunneling parameter generates a boundary crossover scale $T_B$, or equivalently a boundary rapidity $\theta_B\simeq\ln(T_B/T)$. Since the boundary operator has scaling dimension $g$, the crossover scale increases with tunneling amplitude as
\begin{equation}
    T_B\propto |\lambda|^{1/(1-g)}
    \label{eq:end_TB_scaling}
\end{equation}
Note that here $\lambda$ denotes the backscattering strength.
In a quantum-dot-transport device, the transmission through the barrier reads~\cite{kane1992barriers}
\begin{equation}
\mathcal{T}_0=\frac{4\Gamma_t\Gamma_e}{(\Gamma_t+\Gamma_e)^2}.
\end{equation}
As a result,
\begin{equation}
\lambda\propto\sqrt{\mathcal{R}_0}=\sqrt{1-\mathcal{T}_0}=\left|\frac{\Gamma_t-\Gamma_e}{\Gamma_t+\Gamma_e}\right|.
\end{equation}
If $\Gamma_t>\Gamma_e$, in the tunneling regime, lowering the STM tip increases the microscopic tip-sample hopping amplitude $t_0$, and hence increases the bare boundary coupling $\lambda$. As a result, we have Eq.~\eqref{eq:end_TB_scaling}.

The nonperturbative environmental response is evaluated using the rational-string TBA of Ref.~[41]. For a rational dissipation strength $r=p/q$, the dimensionless pseudoenergies satisfy
\begin{equation}
\epsilon_a(\theta)
=
\delta_{a,s}e^\theta
-
\sum_b\int d\theta'\,
\mathcal K_{ab}(\theta-\theta')
\ln\!\left[
1+e^{\bar\mu_b-\epsilon_b(\theta')}
\right],
\end{equation}
where $a,b$ label the rational-string quasiparticles, $s$ is the energy-carrying node, and $\mathcal K_{ab}=K_{ba}/(2\pi)$ is the effective TBA kernel. The two charged strings $c^\pm$ have
\begin{equation}
\bar\mu_{c^\pm}
=
\pm\frac{q eV}{2k_{\rm B}T},
\qquad
\bar\mu_{a\neq c^\pm}=0,
\qquad
\epsilon_{c^+}
=
\epsilon_{c^-}
\equiv\epsilon_{c}.
\end{equation}
Their occupation functions are
\begin{equation}
f_{\pm}(\theta,V)
=
\left\{
1+\exp\!\left[
\epsilon_{c}(\theta,V)
\mp\frac{qeV}{2T}
\right]
\right\}^{-1}.
\end{equation}
Here we have worked in units where $k_{\rm B}=1$.
The environmental current for this rational anchor is
\begin{equation}\label{eq:end_Ienv}
I_{\rm env}(V)
=
\eta
\frac{q e k_{\rm B}T}{h}
\int d\theta\,
\mathcal T(\theta)
\frac{\partial\epsilon_{c}}{\partial\theta}
\left[f_{+}-f_{-}\right],
\end{equation}
where $\eta=\pm1$ is the TBA sign associated with the last string family and
\begin{equation}
\mathcal T(\theta)
=
\frac{1}{
1+\exp\!\left[2p(\theta_B-\theta)\right]},
\qquad
\theta_B=\ln\!\left(\frac{T_B}{T}\right),
\end{equation}
is the exact boundary transmission probability. For a general fitted $r$, the current is obtained from the neighboring rational-anchor currents, and the universal constant-DOS conductance is
\begin{equation}
G_{\rm env}(V;T_B,r,T)
=
\frac{d I_{\rm env}(V;T_B,r,T)}{dV}.
\end{equation}

In the weak-transmission limit this reduces to the usual DCB power law $G_{\rm env}(V,T=0)\propto |eV|^{2r}$, while Eq.~\eqref{eq:end_Ienv} remains valid beyond the perturbative $P(E)$ regime.

\section{STM spectroscopy and smooth approximation}\label{app_smooth_approx}
For STM spectroscopy this environmental response must be combined with the sample spectral weight. Let
\begin{equation}
    \hat\rho(E,\mathbf r_0)
    =
    -\frac{1}{\pi}{\rm Im}\,\hat G^R(E,\mathbf r_0)
\end{equation}
be the local spectral-density matrix of the sample at the tip position. The tip measures the projected DOS
\begin{equation}
    \rho_{\rm tip}(E)=
    \mathbf w^\dagger \hat\rho(E,\mathbf r_0)\mathbf w,
    \label{eq:end_projected_DOS}
\end{equation}
where $\mathbf w$ contains the local tunneling amplitudes into the relevant sample orbitals or bands. In the absence of the dissipative environment, the standard STM expression is
\begin{equation}
    G_0(V)\propto
    \int dE\,\rho_{\rm tip}(E)
    \left[-\partial_E f(E-eV)\right],
    \label{eq:end_plain_STM}
\end{equation}
which gives $G_0(V)\propto \rho_{\rm tip}(eV)$ at low temperature.

With the environment included, the local STM response can be written in terms of an energy-resolved effect described by $\mathcal K_{\rm env}(V,E)$,
\begin{equation}
    G(V)
    =
    \int dE\,\rho_{\rm tip}(E)\,
    \mathcal K_{\rm env}(V,E;T_B,r,T).
    \label{eq:end_energy_kernel}
\end{equation}
This function should satisfy that the constant-DOS limit reproduces the TBA result,
\begin{equation}
    \int dE\,
    \mathcal K_{\rm env}(V,E;T_B,r,T)
    =
    G_{\rm env}(V;T_B,r,T),
    \label{eq:end_kernel_normalization}
\end{equation}
and so that for $r=0$ it reduces to the ordinary STM thermal window,
\begin{equation}
    \mathcal K_{\rm env}(V,E;T_B,0,T)
    \propto
    -\partial_E f(E-eV).
\end{equation}
In the low-temperature spectra analyzed here, the function $\mathcal K$ is sharply localized in the same on-shell energy window that selects the STM DOS, while its total weight is fixed by Eq.~\eqref{eq:end_kernel_normalization}.
Expanding the projected DOS inside this local window gives
\begin{align}
    \rho_{\rm tip}(E)
    =&
    \rho_{\rm tip}(eV)
    +(E-eV)\partial_E\rho_{\rm tip}|_{eV}
    \nonumber\\
    &+
    \frac{(E-eV)^2}{2}\partial_E^2\rho_{\rm tip}|_{eV}
    +\cdots .
\end{align}
Since $\mathcal K$ should be approximately particle-hole symmetric about the on-shell energy, the odd moment is small and the leading relative correction to the product approximation is
\begin{equation}
    \frac{\delta G}{G}
    \sim
    \frac{\sigma_{\rm env}^2}{2}
    \frac{\partial_E^2\rho_{\rm tip}}{\rho_{\rm tip}},
    \label{eq:end_product_error}
\end{equation}
where $\sigma_{\rm env}$ is the effective energy width of the environmental effect. Keeping the leading term gives the product form used in the fits,
\begin{equation}
    G(V)\propto
    \rho_{\rm tip}(eV)\,
    G_{\rm env}(V;T_B,r,T).
    \label{eq:end_DOS_env_product}
\end{equation}
For the vortex curvature analysis, the relevant intrinsic spectral width is $\omega_0\simeq0.42~\mathrm{meV}$, while
$T\simeq0.034~\mathrm{meV}$ at $400$ mK. The thermal-window estimate of the correction is therefore of order
$(T/\omega_0)^2\simeq7\times10^{-3}$. This is small for the zero-bias curvature argument, although corrections can be larger near sharp superconducting coherence peaks where the DOS varies rapidly. Nevertheless, since the superconductivity is not perfect, the coherent peak is not too sharp, as shown in Fig.~\ref{fig:two_channel}, therefore the behavior of superconducting gap can be approximately captured.

\section{Fitting models and protocol}
\label{app_fit_models_strategy}

All spectra were analyzed within the same dissipative STM framework.
For a local spectral component with projected density of states
$\rho(E)$, the measured differential conductance is approximated by
\begin{equation}
    G(V)\simeq \rho(eV)
    G_{\rm env}(V;T_B,r,T_{\rm eff}),
    \label{eq:general-fitting-form}
\end{equation}
where $G_{\rm env}$ is the universal TBA conductance of a
structureless tunneling channel in an Ohmic environment.
The effective temperature $T_{\rm eff}$ accounts for the energy
broadening entering the TBA response. The boundary scale is controlled
by the backscattering strength,
\begin{equation}
    T_{B,a}\propto |\lambda_a|^{(1+r)/r},
    \label{eq:TB-channel}
\end{equation}
where $a$ labels a local tunneling channel.

Before fitting, each spectrum $G_i(V)$ was normalized by its
conductance at $V_{\rm norm}=5$ mV,
\begin{equation}
    \widetilde{G}_i(V)
    =
    \frac{G_i(V)}{G_i(V_{\rm norm})}.
    \label{eq:normalized-conductance}
\end{equation}
All spectra within one setpoint series were fitted globally. The
intrinsic spectral parameters and the environmental parameters were
shared by the full series, whereas the backscattering strength
$\lambda_{i}$ was allowed to vary for each setpoint $i$. The sequence of
$\lambda_{i}$ was constrained to increase (decrease) monotonically with setpoint
current for the first (second) channel, as stated in the maintext.

\subsection{Single-channel vortex fit}
\label{app_single_channel_vortex_fit}

We first fitted the low-energy vortex response using a minimal
single-channel model. The vortex-core density of states was represented
by a broad zero-energy continuum,
\begin{equation}
    \rho_{\rm v}(E)
    =
    \exp\left[
    -\left(\frac{|E|}{\omega_0}\right)^2
    \right]
    +b_{\rm v},
    \label{eq:vortex-dos-single-channel}
\end{equation}
where $\omega_0$ is the intrinsic width of the vortex-core feature and
$b_{\rm v}$ is a smooth background. The fitted conductance is
\begin{equation}
    G_{{\rm v},i}^{(1)}(V)
    =
    \rho_{\rm v}(eV)
    G_{\rm env}
    \left(
    V;T_{B,{\rm v},i},r_{\rm v},T_{\rm eff}
    \right)/G_{\rm norm}.
    \label{eq:vortex-single-channel-fit}
\end{equation}
$G_{\rm norm}$ represents a normalization factor.

The parameters
\begin{equation}
    (r_{\rm v},T_{\rm eff},\omega_0,b_{\rm v})
\end{equation}
were shared by the complete vortex setpoint series. The only
setpoint-dependent quantity was $\lambda_{i}$, which determines
$T_{B,{\rm v},i}$ through Eq.~\eqref{eq:TB-channel}. In particular,
the intrinsic vortex spectral width $\omega_0$ was not allowed to
change with tip height. This fit was designed to determine the
low-energy peak-to-dip boundary flow and was not intended to reproduce
the higher-bias shoulder structure of the full vortex spectra.

\subsection{Two-channel superconducting-gap fit}
\label{app_sc_gap_fit}

The superconducting density of states of each channel was described by
a Dynes form,
\begin{equation}
    \rho_{{\rm SC},a}(E)
    =
    {\rm Re}
    \left[
    \frac{|E|+i\Gamma_a}
    {\sqrt{(|E|+i\Gamma_a)^2-\Delta_a^2}}
    \right],
    \label{eq:dynes-dos}
\end{equation}
where $\Delta_a$ and $\Gamma_a$ are the gap and broadening parameters,
respectively. The full superconducting spectrum was fitted by two
separately dressed channels,
\begin{align}
    G_{{\rm SC},i}^{(2)}(V)
    ={}&
    [\rho_{{\rm SC},1}(eV)
    G_{\rm env}
    \left(
    V;T_{B,1,i},r_{\rm SC},T_{\rm eff}
    \right)
    \nonumber+
    \rho_{{\rm SC},2}(eV)
    G_{\rm env}
    \left(
    V;T_{B,2,i},r_{\rm SC},T_{\rm eff}
    \right)]/G_{\rm norm}.
    \label{eq:two-channel-sc-fit}
\end{align}
We assume that the first (second) channel is in the overcoupled (undercoupled) regime. Therefore, the two superconducting components have identical dissipative environment but parametrically different boundary crossover
scales.

The global fitting parameters were
\begin{equation}
    (\Delta_1,\Delta_2,\Gamma_1,\Gamma_2,
    r_{\rm SC},T_{\rm eff}),
\end{equation}
and $\lambda_{1,i}$ varied between setpoints.

\subsection{Two-channel vortex fit}
\label{app_two_channel_vortex_fit}

The residual shoulder structure in the full vortex spectra was modeled
by adding a weakly projected superconducting contribution to the
strongly accessed vortex-core channel,
\begin{align}
    G_{{\rm v},i}^{(2)}(V)
    ={}&
    [\rho_{\rm v}(eV)
    G_{\rm env}
    \left(
    V;T_{B,1,i},r_{\rm v},T_{\rm eff}
    \right)
    \nonumber+
    \bar{\rho}_{\rm SC}(eV)
    G_{\rm env}
    \left(
    V;T_{B,2,i},r_{\rm v},T_{\rm eff}
    \right)]/G_{\rm norm},
    \label{eq:two-channel-vortex-fit}
\end{align}
where
\begin{equation}
    \bar{\rho}_{\rm SC}(E)
    =
    {\rm Re}
    \left[
    \frac{|E|+i\Gamma_{\rm v,SC}}
    {\sqrt{(|E|+i\Gamma_{\rm v,SC})^2-\Delta_{\rm v,SC}^2}}
    \right].
    \label{eq:residual-sc-vortex-dos}
\end{equation}
Though an SC channel is introduced, the peak-to-dip
evolution at zero bias remains controlled by the dissipative flow of
the vortex-core channel already identified in the single-channel fit.

\subsection{Single-channel high-field fit}
\label{app_single_channel_high_field_fit}
At $B=1$ T, superconductivity is suppressed and the measured spectra
are described by a single metallic channel dressed by the dissipative
environment. To retain the weak particle--hole asymmetry of the normal-state
background, we first extracted a fixed, slowly varying density of states from
the minimum-setpoint spectrum. Specifically, the density of state was fitted by the linear form
\begin{equation}
    \rho_{\rm HF}(E)=aE+b.
    \label{eq:high-field-linear-dos}
\end{equation}
The central region was excluded to prevent the zero-bias DCB anomaly from
being absorbed into the background density of states, while the outermost
bias points were excluded to reduce endpoint effects.

\begin{figure*}[t]
    \centering
    \includegraphics[width=0.5\linewidth]{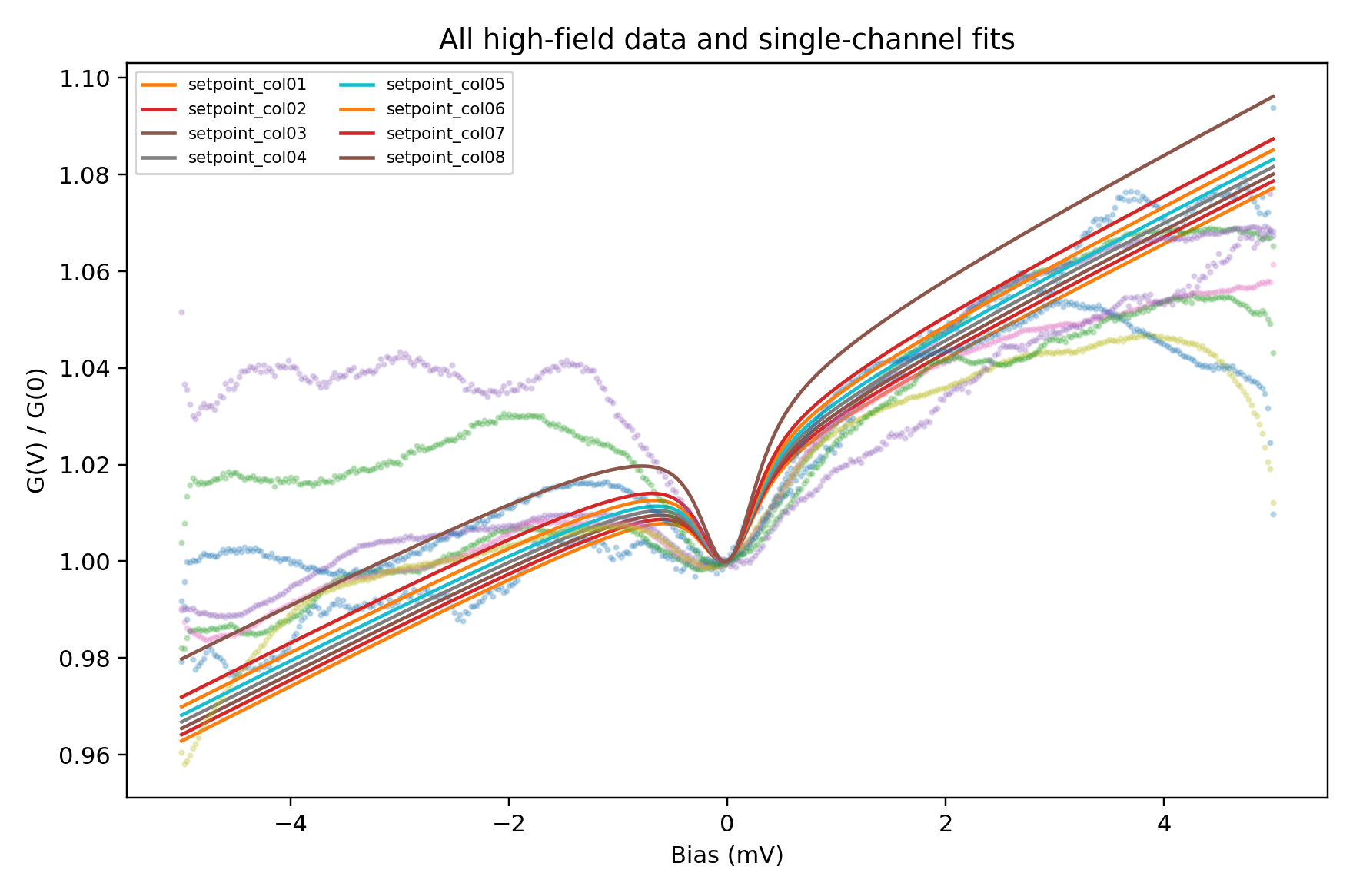}
    \caption{
    The experimental data and corresponding fitting of high-field conductance spectra. The corresponding setpoint current increases from ``setpoint\_col01'' to ``setpoint\_col08''.}
    \label{fig:high-field-fit}
\end{figure*}

For setpoint $i$, the fitted high-field conductance is
\begin{equation}
    G_{{\rm HF},i}(V)
    =
    \rho_{\rm HF}(eV)
    G_{\rm env}
    \left(
        V;T_{B,i},r_{\rm HF},T_{\rm eff}^{\rm HF}
    \right)/G_{\rm norm},
    \label{eq:single-channel-high-field-fit}
\end{equation}
where $G_{\rm env}$ is calculated using the same exact
boundary-sine-Gordon/TBA transport model as in the superconducting and
vortex analyses. The fitted dissipation strength is $r_{\rm HF}\approx0.22$, consistent with the SC-gap and vortex-center cases. The corresponding results are shown in Fig.~\ref{fig:high-field-fit}.

\subsection{Residual weights of fitting}
\label{app_fit_weight_allocation}
\begin{figure*}[t]
    \centering
    \includegraphics[width=0.7\linewidth]{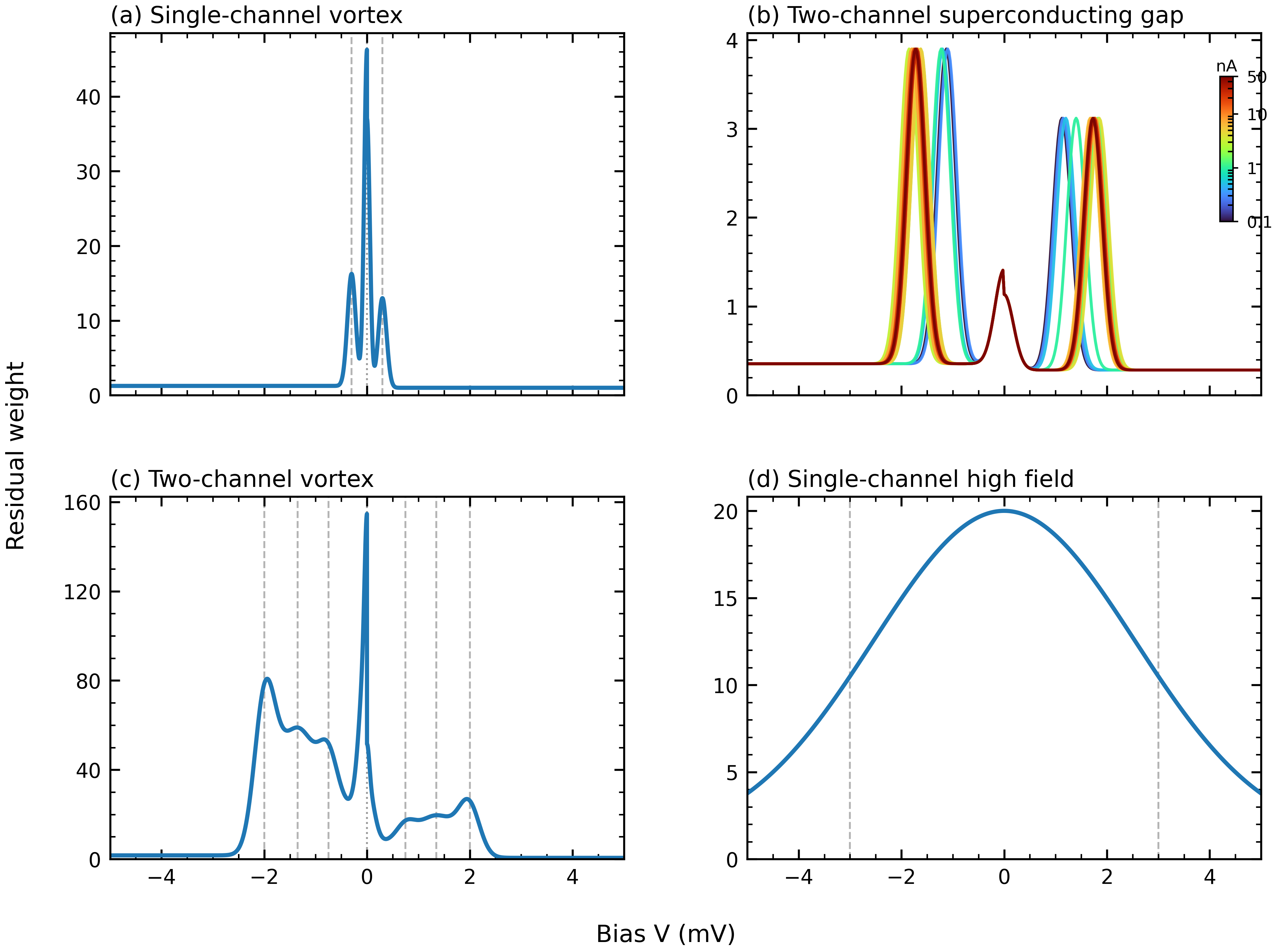}
    \caption{
    Residual-weight profiles used in the global fits.
    (a) Single-channel vortex fit, emphasizing the zero-bias response
    and the reference conductance near $|V|=0.3$ mV.
    (b) Two-channel superconducting-gap fit, with enhanced weights at
    the setpoint-dependent coherence-peak positions and at zero bias.
    (c) Two-channel vortex fit, emphasizing the central vortex
    response, intermediate-bias structure, and superconducting
    shoulders. (d) Single-channel high-field fit, focusing on the low-bias behavior. Dashed lines indicate the reference energies used to
    construct the corresponding weight profiles.
    We have slightly suppress the weight for positive-bias to reduce the influence of noise there, except for the high-field case.
    }
    \label{fig:fitting-weights}
\end{figure*}

For each normalized spectrum, we define the pointwise residual
\begin{equation}
    \delta g_i(V_j)
    =
    \widetilde{G}^{\rm fit}_i(V_j)
    -
    \widetilde{G}^{\rm exp}_i(V_j),
    \label{eq:pointwise-residual}
\end{equation}
where $i$ labels the setpoint spectrum and $V_j$ labels the measured
bias points. The pointwise contribution to the fitting objective is
defined as
\begin{equation}
    \mathcal{L}_{\rm curve}
    =
    \sum_i\sum_j
    W_i(V_j)
    \left[\delta g_i(V_j)\right]^2.
    \label{eq:weighted-residual-loss}
\end{equation}
Equivalently, the residual vector supplied to the numerical optimizer
is
\begin{equation}
    R_{i,j}
    =
    \sqrt{W_i(V_j)}
    \left[
    \widetilde{G}^{\rm fit}_i(V_j)
    -
    \widetilde{G}^{\rm exp}_i(V_j)
    \right].
    \label{eq:weighted-residual-vector}
\end{equation}
Thus, the curves plotted in Fig.~\ref{fig:fitting-weights} show the
multiplicative weight $W_i(V)$ entering the squared residual.

For compactness, we define a Gaussian weight centered at $V_c$,
\begin{equation}
    \mathcal{G}(V;V_c,\sigma)
    =
    \exp\left[
    -\frac{(V-V_c)^2}{2\sigma^2}
    \right],
    \label{eq:gaussian-weight}
\end{equation}
and a smooth inner-bias window,
\begin{equation}
    \mathcal{S}(V;V_{\rm in},\delta V)
    =
    \frac{1}
    {1+\exp\left[(|V|-V_{\rm in})/\delta V\right]}.
    \label{eq:soft-bias-window}
\end{equation}
All weight profiles were fixed before optimization and were not
treated as fitting parameters. A factor $\gamma_-(V)$ was included to give a moderately larger relative weight to the
negative-bias side of the spectra.

For the single-channel vortex fit, greater weight was assigned to the
low-bias region. In particular, the conductance at zero bias and near
$|V|\simeq 0.3$ mV was emphasized so that the fit captures both the
zero-bias response and the peak-to-dip evolution.

For the two-channel superconducting-gap fit, the fitting weights
emphasize the zero-bias suppression and the setpoint-dependent
coherence peaks. This allows the fit to reproduce both the low-energy
DCB feature and the evolution of the superconducting gap structure.

For the two-channel vortex fit, priority was given to the central
vortex feature and the finite-bias structure within the superconducting
energy range. The zero-bias response, the surrounding minima, and the
residual superconducting shoulders were therefore fitted together to
separate the contributions from the two transport channels.

For the single-channel high-field fit, the broad low-bias DCB
suppression was given greater weight, while the full measured bias
range was retained to constrain the slowly varying normal-state
background.

The weights are shown
explicitly in Fig.~\ref{fig:fitting-weights}. These weights serve only to balance spectral regions with distinct physical roles; the
intrinsic spectral and environmental parameters are determined from the global fit to the full setpoint series.

\section{Summary of fitted parameters}
\label{app_fit_parameter_summary}

For completeness, we summarize the optimized parameters of the three
global fits introduced above. Within each setpoint series, the intrinsic
spectral parameters and the environmental parameters were shared by all
spectra, whereas the backscattering strength $\lambda_{i}$ was optimized
separately for each setpoint $i$ under the monotonicity constraint described
above.

Table~\ref{tab:global_fit_parameters} lists the physically relevant global
parameters. The background parameter $b_v$ depends on
the normalization and on the number of spectral components included in the
model; it should therefore be regarded as a nuisance parameter rather than
an intrinsic spectral observable.

\begin{table*}[t]
\caption{
Summary of the physically relevant global parameters obtained from the
four fitting procedures. All entries are optimized fitting parameters
unless marked by a dagger. For the high-field fit, the daggered
coefficients define the linear density of states
$\rho_{\rm HF}(E)=a_{\rm HF}E+b_{\rm HF}$; they were determined from
the minimum-setpoint spectrum and subsequently held fixed. A dash
indicates that the corresponding parameter is absent from the model.
}
\label{tab:global_fit_parameters}
\begin{ruledtabular}
\begin{tabular}{lcccc}
Parameter
& \shortstack{SC gap\\two channel}
& \shortstack{Vortex\\one channel}
& \shortstack{Vortex\\two channel}
& \shortstack{High field\\one channel} \\
\hline

$r$
& $0.20596$
& $0.19294$
& $0.18108$
& $0.22222$ \\

$T_{\mathrm{eff}}$ (K)
& $0.51466$
& $0.92741$
& $1.89275$
& $0.83406$ \\

$\Delta_{1}$ (meV)
& $0.75573$
& ---
& ---
& --- \\

$\Gamma_{1}$ (meV)
& $0.32021$
& ---
& ---
& --- \\

$\Delta_{2}$ (meV)
& $1.43537$
& ---
& ---
& --- \\

$\Gamma_{2}$ (meV)
& $0.28585$
& ---
& ---
& --- \\

$\omega_{0}$ (meV)
& ---
& $0.34177$
& $0.50626$
& --- \\

$b_{\rm v}$
& ---
& $3.74750$
& $3.74765$
& --- \\

$\Delta_{\rm v,\mathrm{SC}}$ (meV)
& ---
& ---
& $0.91042$
& --- \\

$\Gamma_{\rm v,\mathrm{SC}}$ (meV)
& ---
& ---
& $0.52512$
& --- \\

$a_{\rm HF}$ (meV$^{-1}$)
& ---
& ---
& ---
& $0.010516^{\dagger}$ \\

$b_{\rm HF}$
& ---
& ---
& ---
& $0.93784^{\dagger}$ \\

\end{tabular}
\end{ruledtabular}
\end{table*}

The three fittings independently give
\begin{equation}
r_{\mathrm{SC}}=0.20596,
\qquad
r_{\rm v}^d=0.18108,
\qquad
r_{\rm HF}=0.22222
\end{equation}
showing that the three data sets are governed by nearly the same dissipative
environment.

\end{document}